\documentclass[reprint,twocolumn,amsmath,amssymb,aps,superscriptaddress]{revtex4}
\usepackage{graphicx}
\usepackage{dcolumn}
\usepackage{color}
\usepackage{pstricks,pst-grad,color}
\usepackage{bm}
\usepackage{ulem}
\usepackage{calc}
\usepackage{braket}
\usepackage{nccmath}
\usepackage{comment}
\usepackage{here}

\providecommand{\U}[1]{\protect\rule{.1in}{.1in}}

\begin{document}
\title{Relationship between exceptional points and the Kondo effect in $f$-electron materials}

\author{Yoshihiro Michishita}
\affiliation{Department of Physics, Kyoto University, Kyoto 606-8502, Japan}
\author{Tsuneya Yoshida}
\affiliation{Department of Physics, Tsukuba University, Tsukuba, Japan}
\author{Robert Peters}
\affiliation{Department of Physics, Kyoto University, Kyoto 606-8502, Japan}

\begin{abstract}
We study the impact of nonhermiticity  due to strong correlations in $f$-electron materials. One of the most remarkable phenomena occurring in nonhermitian systems is the emergence of exceptional points at which the effective nonhermitian Hamiltonian cannot be diagonalized.
We here demonstrate that the temperature at which exceptional points appear around the Fermi energy is related to the Kondo temperature.
For this purpose, we study the periodic Anderson model with local and nonlocal hybridization in the insulating and metallic regimes.
By analyzing the effective nonhermitian Hamiltonian, which describes the single-particle spectral function, and the temperature dependence of the magnetic moment, we show that  exceptional points appear at the temperature at which the magnetic moment is screened. This temperature corresponds to the Kondo temperature.
 These results suggest that the well-known crossover between localized and itinerant $f$-electrons in these materials is related to the emergence of exceptional points in the single-particle spectral function at the Fermi energy. Viewing exceptional points in the combined momentum-frequency space, we observe that the exceptional points in the effective Hamiltonian form a one-dimensional manifold which changes its structure around the Kondo temperature.
\end{abstract}

\newcommand{\rp}[1]{{\color{red} #1}}
\newcommand{\ym}[1]{{\color{red} #1}}
\newcommand{\cm}[1]{}
\newcommand{\cmm}[1]{{\color{red}\sout{ #1}}}

\maketitle
\section{Introduction}
Recently, phenomena described by an effective nonhermitian Hamiltonian are intensively studied especially in the context of artificial quantum systems.\cite{PhysRevLett.77.570,PhysRevB.56.8651, PhysRevLett.108.173901,brandstetter2014reversing, san2016majorana, doppler2016dynamically,PhysRevLett.116.133903,ashida2017parity,chen2017exceptional,feng2017non,PhysRevLett.120.146402, PhysRevX.8.031079,PhysRevLett.121.203001}
Effective nonhermitian Hamiltonian can induce novel topological phases\cite{PhysRevLett.120.146402,PhysRevX.8.031079,PhysRevLett.121.026808,PhysRevLett.121.086803,PhysRevLett.121.136802,kawabata2018symmetry,kawabata2019topological} and novel phenomena such as anomalous edge states\cite{PhysRevLett.116.133903,PhysRevLett.121.086803},  unusual quantum critical phenomena\cite{PhysRevLett.121.203001,ashida2017parity,yamamoto2019theory}, unidirectional invisibility \cite{PhysRevLett.106.213901,regensburger2012parity,feng2014single}, chiral transport \cite{PhysRevLett.86.787,peng2014parity,gao2015observation,doppler2016dynamically,xu2016topological}, and enhanced sensitivity \cite{PhysRevX.4.041001,PhysRevLett.112.203901,PhysRevLett.117.110802,hodaei2017enhanced,chen2017exceptional,yoon2018time,lau2018fundamental}.

In open quantum systems, e.g. in cold atomic systems, it is possible to derive an effective nonhermitian Hamiltonian under certain conditions even though the Hamiltonian describing the total system is hermitian\cite{PhysRevA.85.032111,PhysRevA.94.053615}. 
However, as the system becomes larger, it is difficult to experimentally realize these conditions, such as postselection or a $\mathcal{PT}$-symmetric setup. 
Thus, experiments about nonhermitian phenomena in artificial quantum systems are particularly done in one-dimensional or small systems.\cite{PhysRevLett.103.093902,ruter2010observation,doppler2016dynamically,xu2016topological,hodaei2017enhanced,yoon2018time}

On the other hand, strongly correlated systems in equilibrium have not been considered to be related to nonhermitian systems until recently. Generally, the band structure of strongly correlated materials, which is given by the single-particle spectral function, is renormalized and broadened by the self-energy, $\Sigma(\omega)$. The spectral function in equilibrium can always be written as $A(\omega)=-\frac{1}{\pi}\mathrm{Im}\Bigl\{\mathrm{tr}(\omega-\mathcal{H}_{eff})^{-1}\Bigr\}$, where $\mathcal{H}_{eff}=\mathcal{H}_0 + \Sigma$, $\mathcal{H}_0$ is the non-interacting part of the Hamiltonian, and $\Sigma$ is the self-energy. The imaginary part of the self-energy can be related to the finite life time of the quasi-particles in the strongly correlated material. 
Until recently, the effect of the imaginary part of the self-energy has been merely considered as a broadening of the spectral function. However, Kozii and Fu \cite{kozii2017non} have shown that because of the imaginary part of the self-energy, the effective Hamiltonian describing the spectral function is nonhermitian, which can generate exotic phenomena\cite{PhysRevLett.120.146402,PhysRevLett.121.026403,PhysRevB.98.035141}.
For example, the effective nonhermitian Hamiltonian can be defective at an exceptional point (EP) in the Brillouin zone (BZ), where it cannot be diagonalized. At these EPs, a topological number can be defined\cite{PhysRevLett.120.146402}.
Moreover, different EPs in the BZ might be connected by bulk Fermi arcs which could be observed in ARPES spectroscopy.
These nonhermitian phenomena, which can be seen in the equilibrium state of strongly correlated materials, are now studied vigorously. They also hold the potential to explain the pseudo-gap in cuprate superconductors or quantum oscillations in the topological Kondo insulators\cite{PhysRevLett.121.026403} SmB$_6$\cite{Tan287,liu2018fermi} and  YbB$_{12}$\cite{xiang2018quantum}.

We note that to obtain an effective nonhermitian Hamiltonian describing the spectral function, postselection or other difficult experimental setups are not necessary. Therefore, it seems reasonable 
that 2D or 3D bulk nonhermitian phenomena can be more easily observed in strongly correlated materials than in artificial quantum systems.
  
  It has been shown that the minimal model which can include exceptional points in the spectral function must consist of at least two hybridized bands, which include different self-energy. A model exactly describing this situation is the periodic Anderson model (PAM), which consists of an uncorrelated band which is hybridized with a strongly correlated band.
  This model is generally used to describe $f$-electron materials, where the uncorrelated band describes conduction (c) electrons and the correlated band describes $f$-electrons. Because of the strong correlations, many remarkable phenomena can be observed in $f$-electron materials, such as magnetism, unconventional superconductivity, quantum criticality, and the Kondo effect.
  
  In this paper, we study  nonhermitian phenomena induced by the self-energy in the Kondo regime of 2D $f$-electron materials by using the dynamical mean field theory (DMFT)\cite{RevModPhys.68.13} combined with the numerical renormalization group (NRG)\cite{RevModPhys.47.773,RevModPhys.80.395,PhysRevB.74.245114}. 
  We elucidate the relationship between the appearance of exceptional points (or exceptional loops) at the Fermi energy in the spectral function and the transition from a metal at high temperatures to the Kondo insulator or the heavy fermion state at low temperature. Thus, the appearance of exceptional points at the Fermi energy corresponds to the transition from localized $f$-electrons to itinerant $f$-electrons.
  We note that the emergence of exceptional points and bulk Fermi arcs has been reported in different versions of the periodic Anderson model\cite{YukiNagai} and the Kondo lattice model\cite{PhysRevB.98.035141}.
  
  The rest of this paper is organized as follows. In Sec. \ref{Model}, we introduce the models and we briefly explain about exceptional points and the related nonhermitian topological numbers in strongly correlated electron systems.
  In Sec. \ref{Results}, we show the numerical results by DMFT/NRG about the Kondo temperature and the temperature at which the exceptional points emerge at the Fermi surface. In Sec. \ref{EPR}, we analyze the structure of exceptional points in the combined momentum-frequency-space.  In Sec. \ref{Discuss}, we conclude this paper.

  \section{Models and Non-hermitian properties in SCES}\label{Model}
  To analyze the emergence of exceptional points and the Kondo effect in $f$-electron materials, we use the periodic Anderson model,
 \begin{align}
&{\cal H}\mathalpha{=}\!\sum_{\bm{k}}\Bigl((\epsilon_{\bm{k}}+\mu_c)c^{\dagger}_{\bm{k}\sigma}c_{\bm{k}\sigma}
 \mathalpha{+} (\epsilon_{f\bm{k}}+\mu_f) f^{\dagger}_{\bm{k}\sigma}f_{\bm{k}\sigma}\nonumber\\
& \ \ \ \ \ \ \ \ \ \  \mathalpha{+}(V_{l/p})_{\sigma\sigma^{\prime}}(f^{\dagger}_{\bm{k}\sigma}c_{\bm{k}\sigma^{\prime}}\mathalpha{+}h.c)\Bigr)\mathalpha{+}U \sum_i n_{i\uparrow}n_{i\downarrow}\label{X}\\
& \ \ \epsilon_{c/f} = -2 t_{c/f} (\cos{k_x}+\cos{k_y}) \\
& \ \ V_l = V \delta_{\sigma \sigma^{\prime}} \\
& \ \ V_p = V (\bm{\sigma} \cdot \sin{\bm{k}}) \ \ \ \ \ \bigl(\sin{\bm{k}}= (\sin{k_x},\sin{k_y}) \bigr)\\
& \ \ n_{i\sigma}=f^{\dagger}_{i\sigma}f_{i\sigma}
\end{align}
where $c^{(\dagger)}_{\bm{k}\sigma},f^{(\dagger)}_{\bm{k}\sigma}$ are annihilation (creation) operators of the $c$- and the $f$-electrons for momentum $\bm{k}$ and spin-direction $\sigma$. $t_{c,f}$ are the inter-site hopping strengths for the $c$- and the $f$-electrons. For simplicity, we assume a two-dimensional square lattice.
\begin{figure}[t]
\begin{minipage}[t]{4.2cm}
\includegraphics[width=4.0cm]{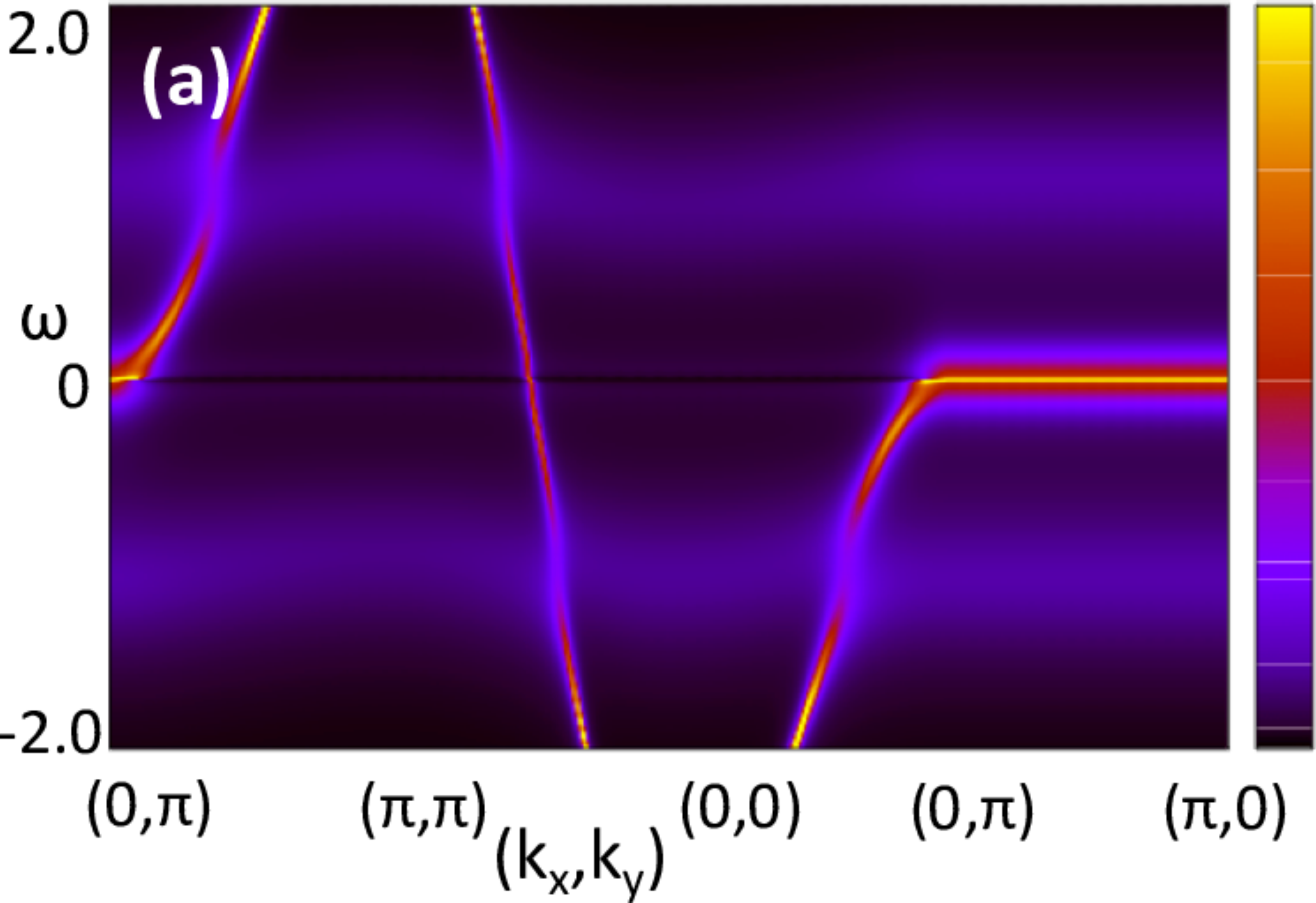}
\end{minipage}
\begin{minipage}[t]{4.2cm}
\includegraphics[width=4.0cm]{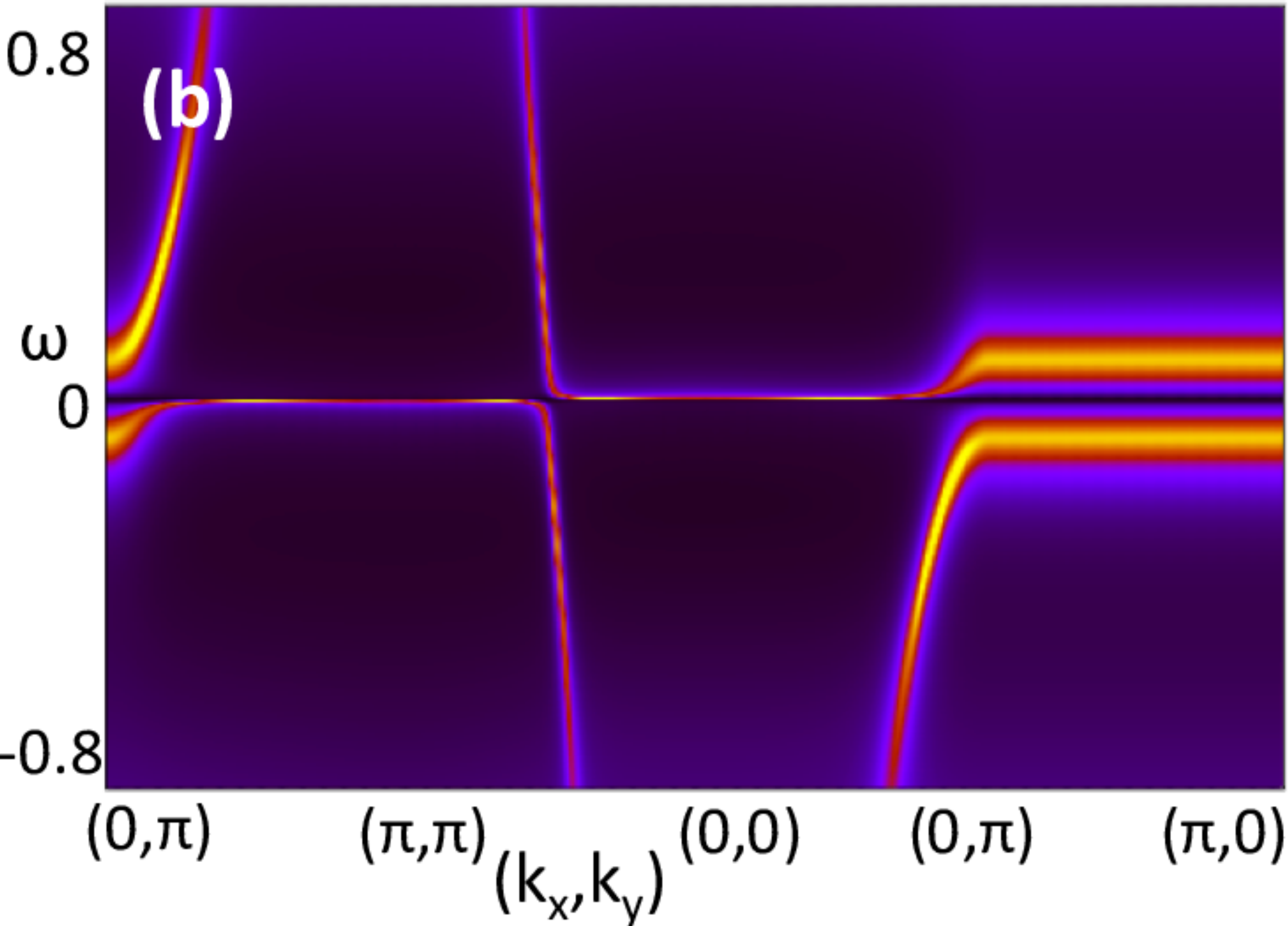}
\end{minipage}
\vspace{0.8cm}
\begin{minipage}[t]{4.2cm}
\includegraphics[width=4.0cm]{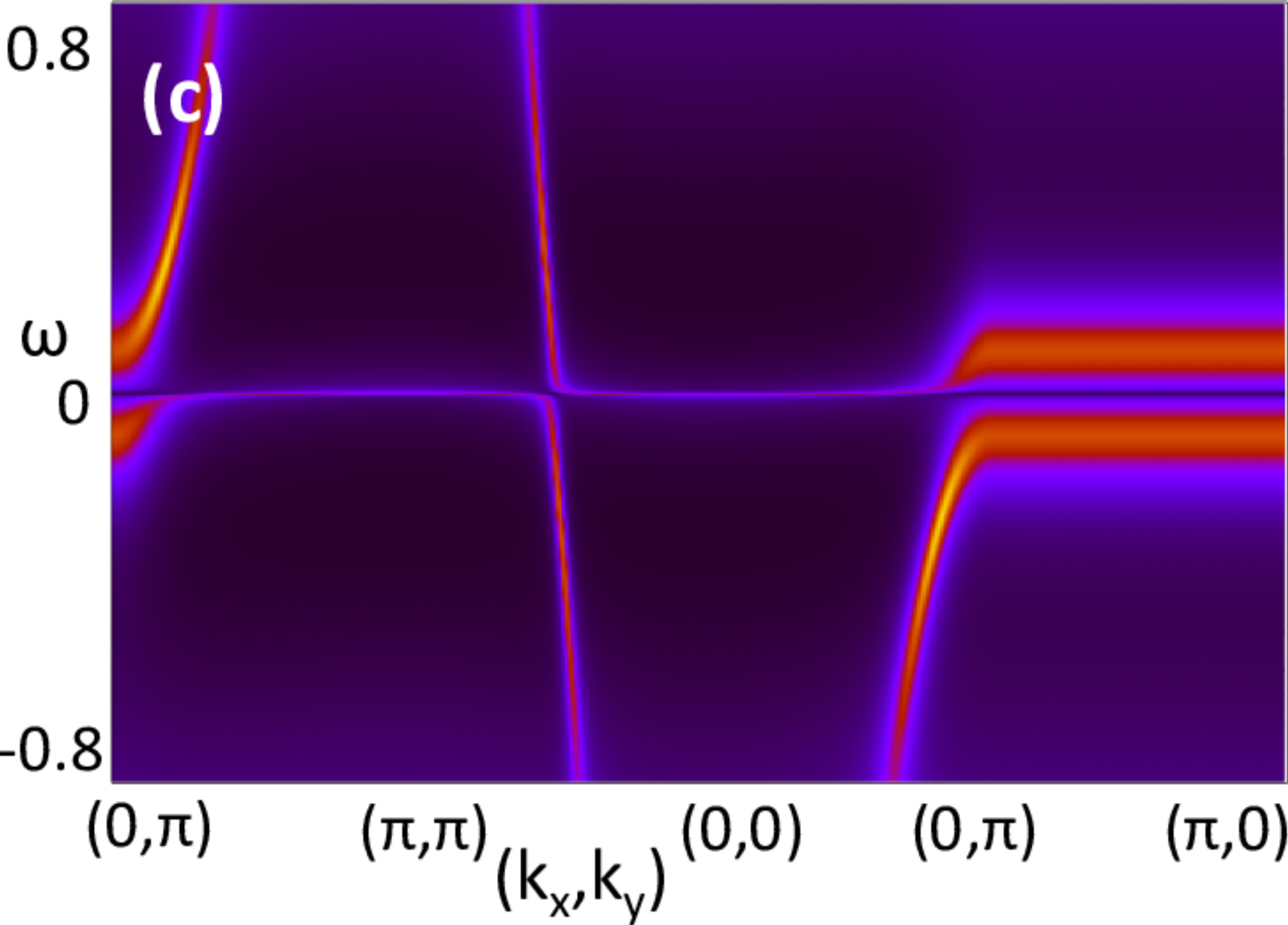}
\end{minipage}
\begin{minipage}[t]{4.2cm}
\includegraphics[width=4.0cm]{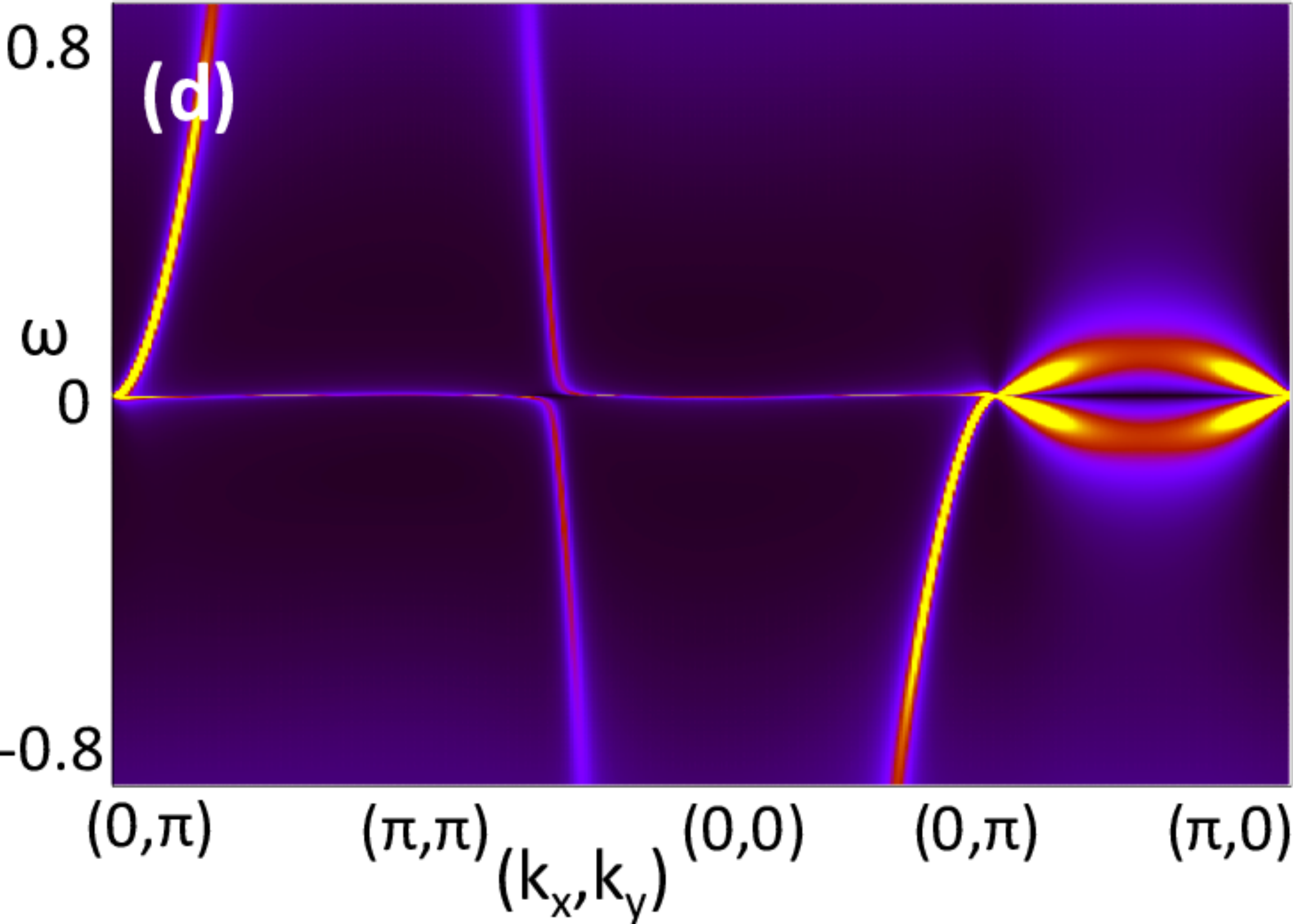}
\end{minipage}
\caption{  (a)-(d) : Momentum-resolved spectral functions for the Kondo insulator, the heavy-fermion state, and the metallic state with p-wave hybridization for  $V$=0.4. Fig.\ref{fig:Disp}(a) shows a high temperature spectral function, $T$=0.13 of the Kondo insulator. Fig.\ref{fig:Disp}(b)-(d) show spectral functions at low temperatures, for $T$=0.0005,  for the Kondo insulator, the heavy-fermion state, and the nonlocal hybridization, respectively. \label{fig:Disp}}
\begin{minipage}[t]{4.2cm}
\includegraphics[width=4.0cm]{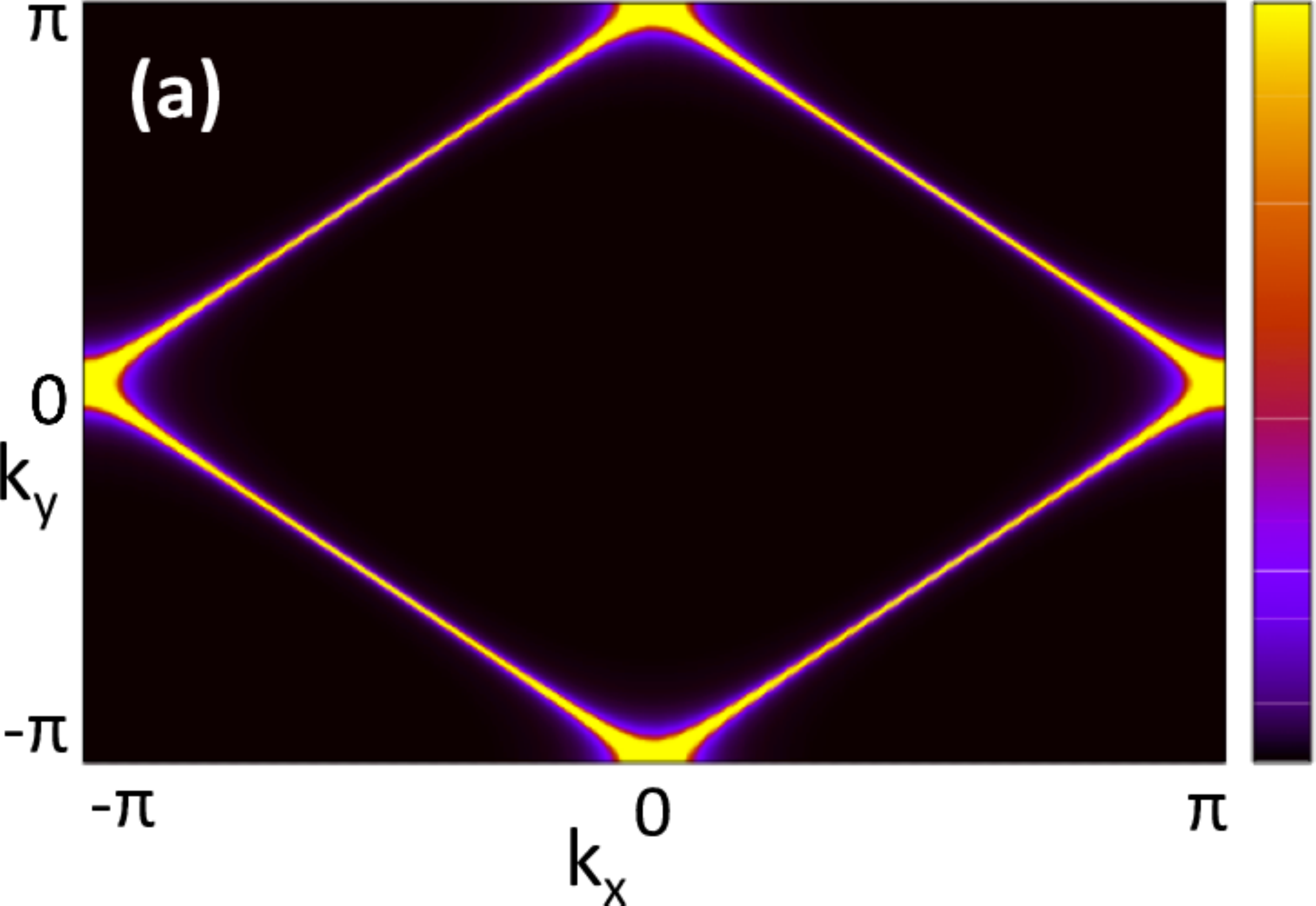}
\end{minipage}
\begin{minipage}[t]{4.2cm}
\includegraphics[width=4.0cm]{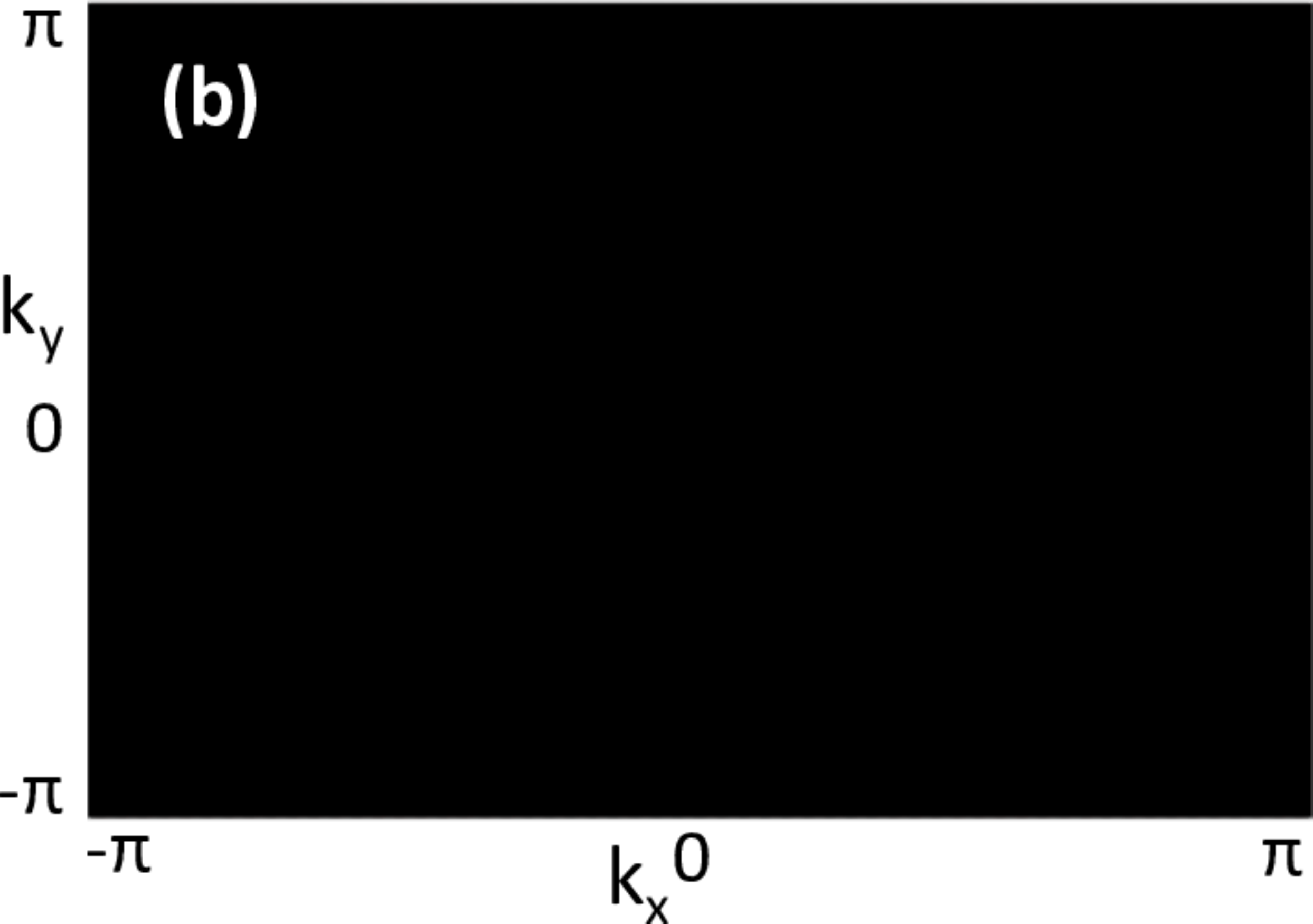}
\end{minipage}
\vspace{0.8cm}
\begin{minipage}[t]{4.2cm}
\includegraphics[width=4.0cm]{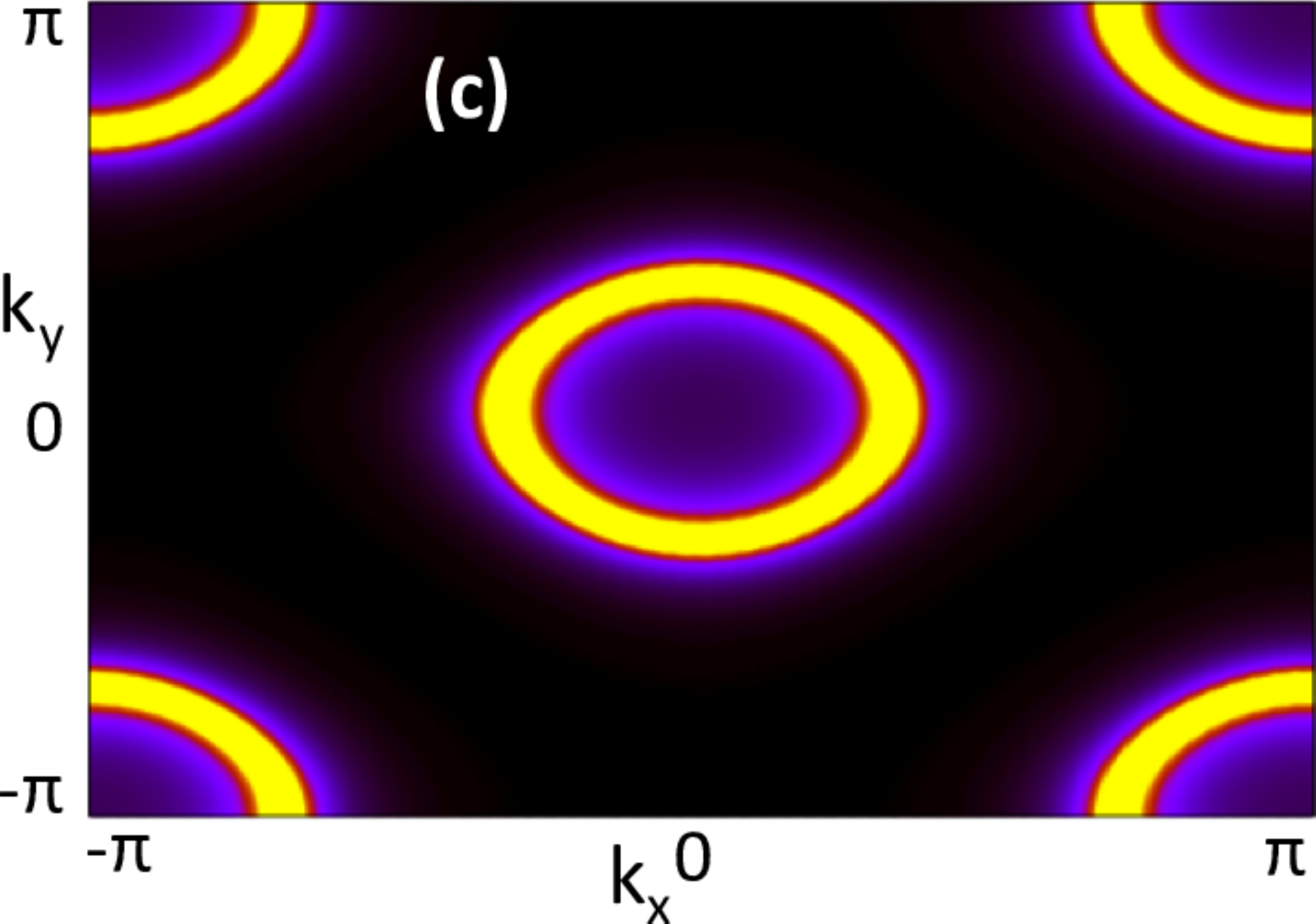}
\end{minipage}
\begin{minipage}[t]{4.2cm}
\includegraphics[width=4.0cm]{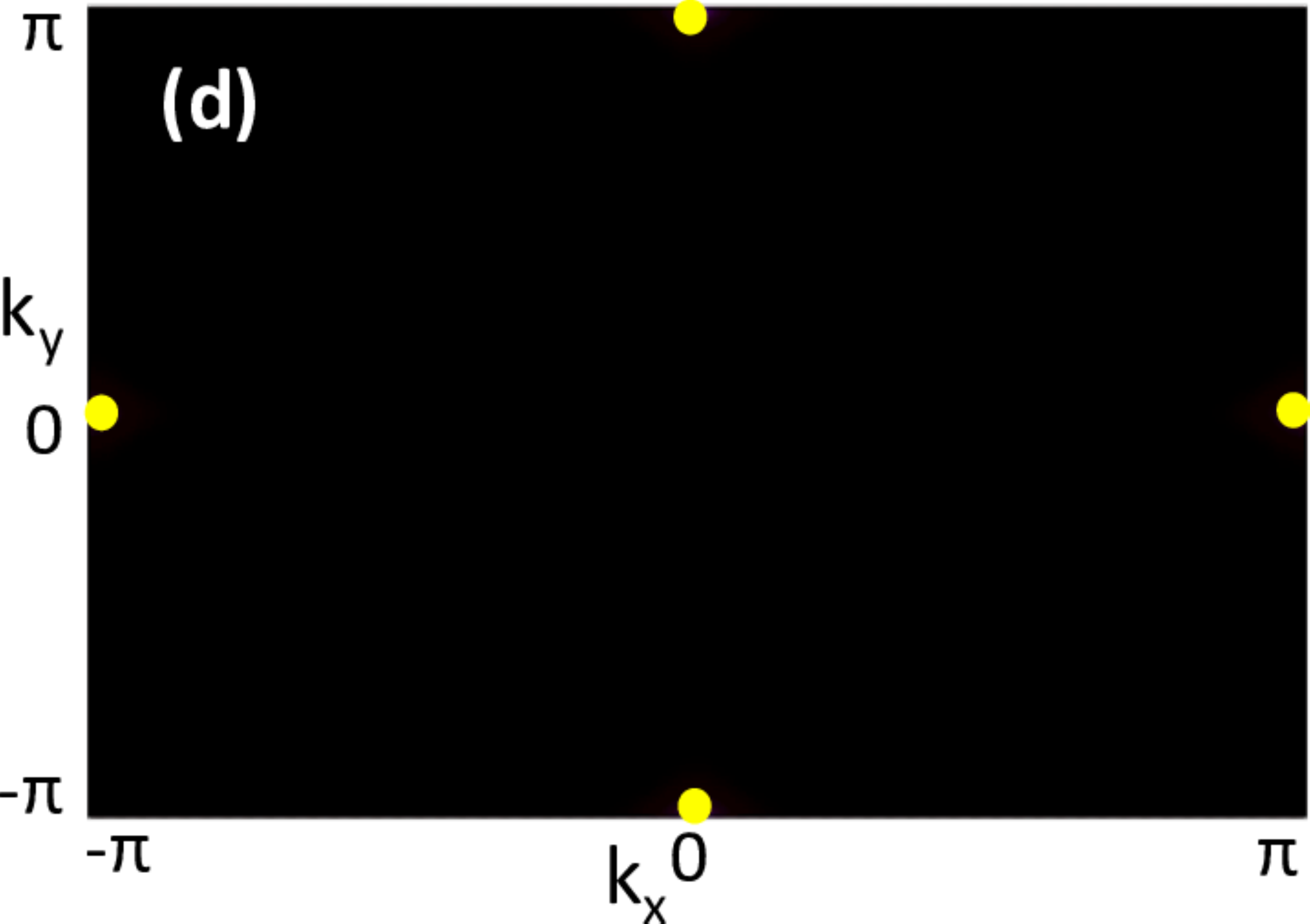}
\end{minipage}
\caption{  (a)-(d) : Momentum-resolved spectral functions at $\omega=0$ corresponding to the spectral functions shown in Fig.\ref{fig:Disp}. In (d), we have enhanced the visibility of the Fermi surface by changing the color.  \label{fig:FS}}
\end{figure}
$\mu_{c/f}$ are the chemical potentials for the $c$- and $f$-orbitals. $V_{l/p}$ describe a local and a nonlocal hybridization between the $c$- and $f$-orbitals, respectively.
Throughout this paper, we fix $t_f= \pm 0.05t_c$, $\mu_c=0$, $\mu_f=-1.0$, $U=2.0$ and use $t_c=0.8$. Using this model, we analyze the relation between the Kondo effect and the emergence of  exceptional points. We will focus on three different cases:  $t_f = -0.05 t_c$ with a local hybridization($V_l\neq0$ ,$V_p=0$), $t_f = 0.05 t_c$  with a local hybridization($V_l\neq0$ ,$V_p=0$), and $t_f = -0.05 t_c$ with a p-wave hybridization($V_l=0$ ,$V_p\neq0$). 

In Figs. \ref{fig:Disp} and \ref{fig:FS}, we show the momentum resolved spectral functions and the Fermi surfaces for all three states.
At high temperature, the $f$-electrons are localized and do not hybridize with the $c$-electrons, as shown in Fig. \ref{fig:Disp}(a) and Fig. \ref{fig:FS}(a). Below the Kondo temperature, $f$-electrons become itinerant and hybridize with the $c$-electrons, which results in strong changes in the spectral function. Figure \ref{fig:Disp}(b) shows the spectral function of the Kondo insulator having a gap at the Fermi energy. Figure \ref{fig:Disp}(c) shows the spectral function of the metallic regime with local hybridization,  and Fig. \ref{fig:Disp}(d) the spectral function of the $p$-wave hybridization.
Corresponding to these spectral functions, we show the spectral weight at the Fermi energy in Fig. \ref{fig:FS}. At high temperatures, Fig. \ref{fig:FS}(a), we only find the $c$ electrons at the Fermi energy. At low temperatures, all three states have very different Fermi surfaces. The Kondo insulating state, shown in Fig. \ref{fig:FS}(b), has no spectral weight at the Fermi energy. The heavy-Fermion state, Fig. \ref{fig:FS}(c), shows the Fermi surface corresponding to the metallic state. Finally, in Fig. \ref{fig:FS}(b), the point-like Fermi surface of the metallic state with $p$-wave hybridization is shown.

We employ the DMFT combined with the NRG to calculate the physical properties in these models. DMFT takes local fluctuations fully into account by self-consistently solving the mean field equations.\cite{RevModPhys.68.13}
The lattice Hamiltonian is thereby mapped onto a quantum impurity model. DMFT neglects nonlocal fluctuations. Even though nonlocal fluctuations might not be small in 2D systems and even crucial for the magnetic state, they might be less important for the Kondo effect and the emergence of exceptional points. Furthermore, all shown results remain correct in three-dimensional systems, where nonlocal fluctuations are weaker compared to the 2D system.
 To solve the quantum impurity model, we use the NRG, which calculates low energy properties by iteratively  discarding high-energy states. It has been shown that NRG is a very reliable tool at low temperature.\cite{RevModPhys.80.395,PhysRevB.74.245114}

Before showing the numerical results, we briefly introduce exceptional points in strongly correlated materials.
 As mentioned above, the periodic Anderson model is one of the minimum model for the emergence of the exceptional points. The effective nonhermitian Hamiltonian which describes the spectral function can be written by
\begin{align}
  &\mathcal{H}_{eff}(\bm{k},\omega)= \mathcal{H}_0 + \Sigma(\omega) = \ \left(
    \begin{array}{cc}
      \epsilon_c(\bm{k}) & V(\bm{k}) \\
      V(\bm{k}) & \epsilon_f(\bm{k}) + \Sigma(\omega) \\
    \end{array}
  \right)\nonumber\\
  & \ \ \ \ \ \ \ \ \ \ \ = \ \  h_0 \bm{1} + h_1 \sigma^z + V(\bm{k}) \sigma^x\\
  \nonumber\\
  & \ \ h_0 = \ \  (\epsilon_c(\bm{k}) + \epsilon_f(\bm{k}) + \Sigma(\omega))/2\\
  & \ \ h_1 = \ \  (\epsilon_c(\bm{k}) - \epsilon_f(\bm{k}) - \Sigma(\omega))/2\\
    \nonumber\\
  &E_{\pm}-h_0 =  \pm \sqrt{h_1^2 + V^2(\bm{k})} \\
  &= \mathalpha{\pm}\Biggl\{\left(\frac{\left(\epsilon_c(\bm{k})\mathalpha{-}\epsilon_f(\bm{k})\mathalpha{-}\mathrm{Re}\Sigma(\omega)\right)^2}{4}\mathalpha{+}V^2(\bm{k})\mathalpha{-}\frac{\left(\mathrm{Im} \Sigma(\omega)\right)^2}{4} \right)\nonumber\\
  & \ \ \ \ \ \ \ \ \ \ \mathalpha{+}\frac{i}{2}\biggl( \mathrm{Im} \Sigma(\omega)\Bigl(\epsilon_c(\bm{k})\mathalpha{-}\epsilon_f(\bm{k})\mathalpha{-}\mathrm{Re}\Sigma(\omega)\Bigr)  \biggr)\Biggr\}^{\frac{1}{2}},
\end{align}
 where $E_{\pm}$ are the eigenvalues of the effective Hamiltonian. For the system with $p$-wave hybridization, we use the helical basis, in which $V(\bm{k})=V_p\sqrt{\sin^2(k_x)+\sin^2(k_y)}$.
 This effective nonhermitian Hamiltonian becomes nondiagonalizable when the following conditions are satisfied:
\begin{align}
& \epsilon_c(\bm{k}) - \epsilon_f(\bm{k}) - \mathrm{Re}\Sigma(\omega) =0\label{1}\\
& \mathrm{Im} \Sigma(\omega) /2 = V(\bm{k}) .\label{2}
  \end{align}
  These points (sometimes loops) in the momentum space, for which the nonhermitian Hamiltonian cannot be diagonalized, are called ”exceptional points.” Moreover, we can define a winding number on these points which reads\cite{PhysRevLett.120.146402},
\begin{align}
W = \oint_{\mathrm{EP}} \frac{d\bm{k}}{2\pi i} \cdot \bm{\nabla}_{\bm{k}} \mathrm{log} \  \mathrm{det} \mathcal{H}_{eff} (\bm{k},\omega).
\end{align}

  Exceptional points with $W\neq 0$ are topologically stable because $W$ does not change unless the exceptional point is annihilated with another one. 
  We note that, in strongly correlated materials, the effective nonhermitian Hamiltonian is introduced for describing the spectral function\cite{PhysRevB.98.035141}. Therefore, when $(\omega-h_0)$ is not small, the spectral weight at the exceptional points is small and might only have a little effect on  observable phenomena. We will thus distinguish exceptional points with $\mathrm{Re}(\omega-h_0)\simeq0$ from the exceptional points where $|\mathrm{Re}(\omega-h_0)|$ is large. In this paper, we call the former "exceptional points (EPs)" and the later "irrelevant exceptional point (iEP)." In short, iEPs have less spectral weight and therefore are less relevant to physical phenomena than EPs.
  
\begin{figure}[t]
\begin{minipage}[t]{4.2cm}
\includegraphics[width=4.0cm]{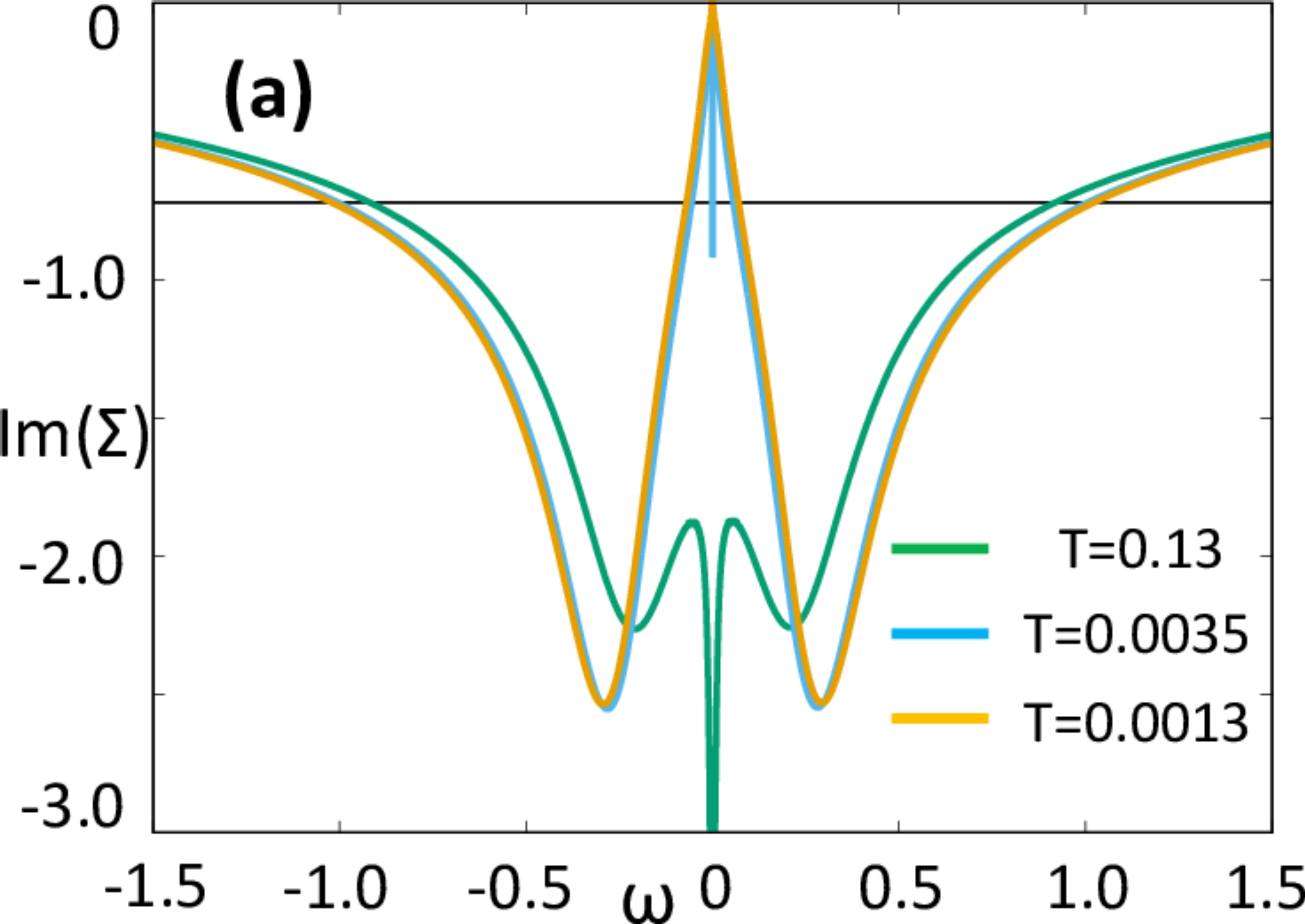}
\end{minipage}
\begin{minipage}[t]{4.2cm}
\includegraphics[width=4.0cm]{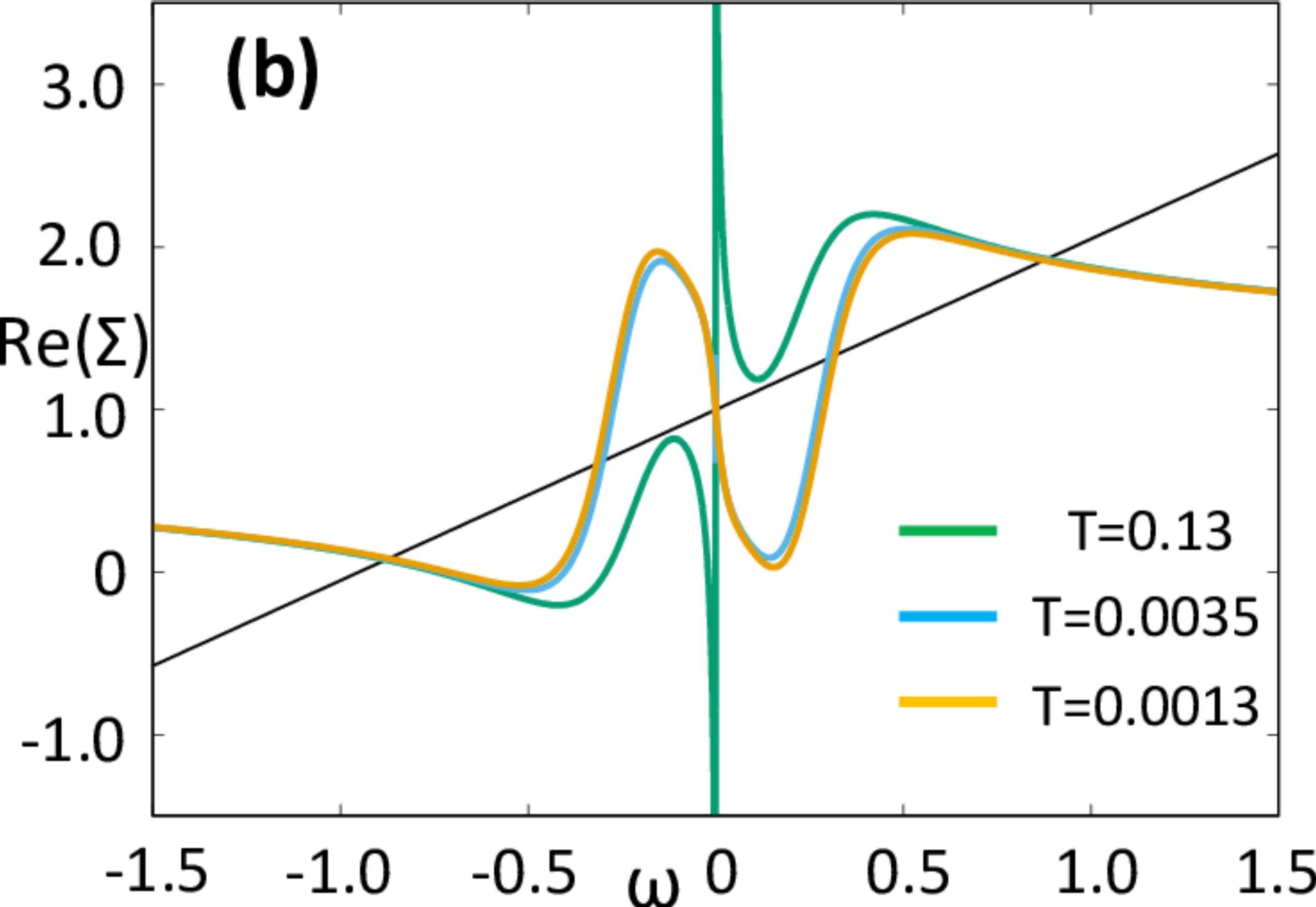}
\end{minipage}
\vspace{0.8cm}
\begin{minipage}[t]{4.2cm}
\includegraphics[width=4.0cm]{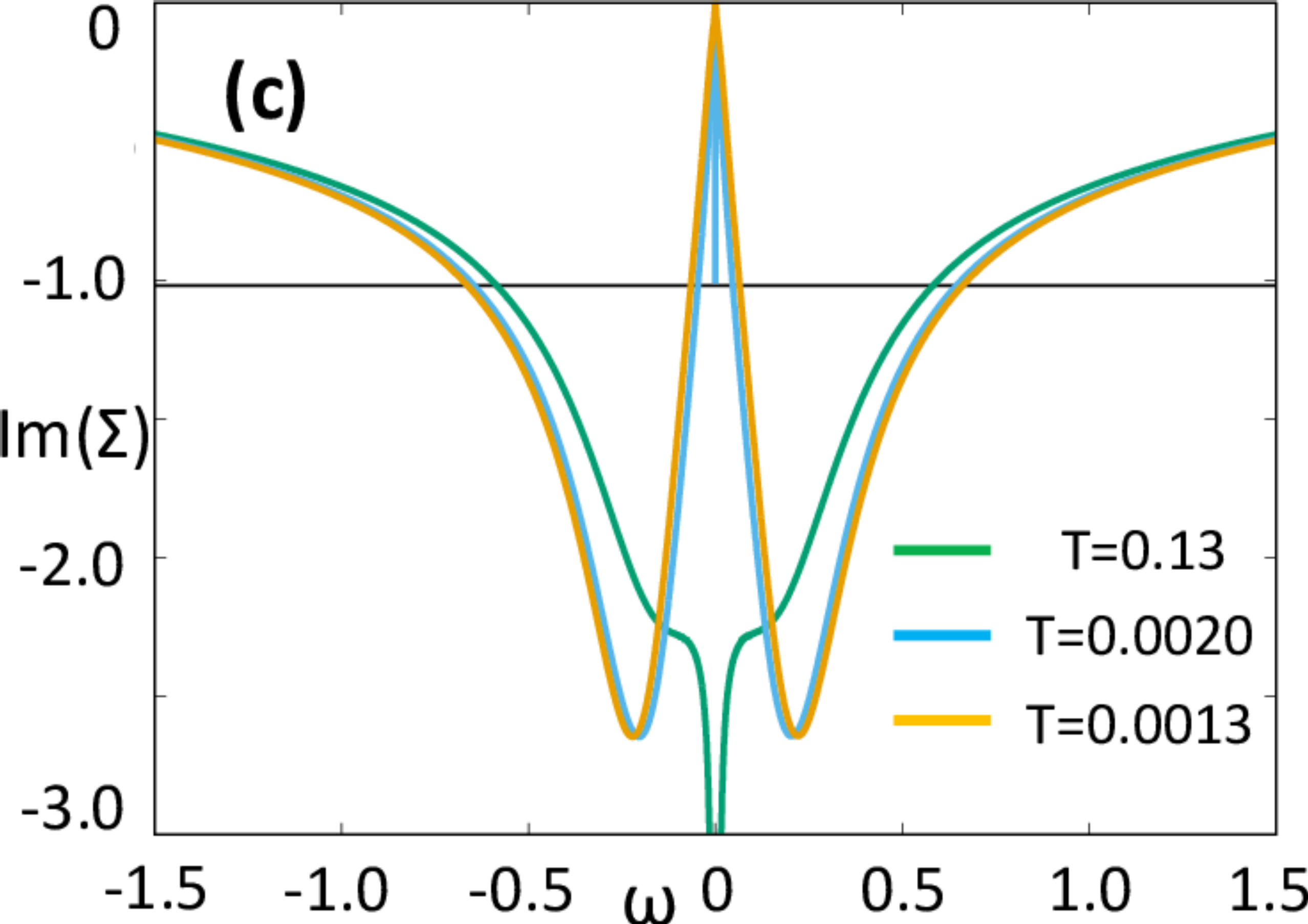}
\end{minipage}
\begin{minipage}[t]{4.2cm}
\includegraphics[width=4.0cm]{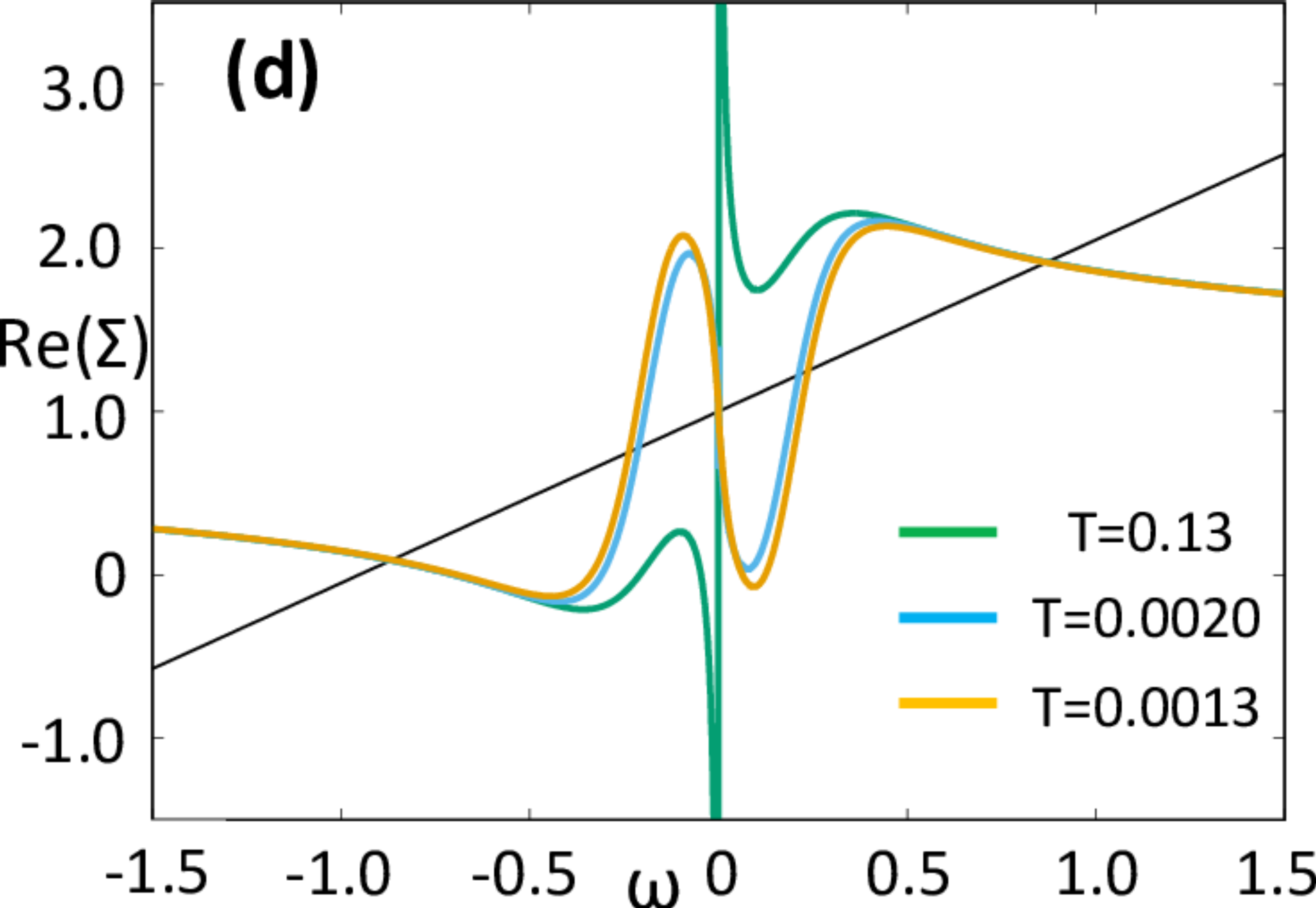}
\end{minipage}
\caption{  (a)-(d) : The temperature dependence of the imaginary and the real part of the self-energy calculated by DMFT/NRG.
(a) and (b) show the results for the local hybridization with $V$=0.36 and $t_f=-0.05 t_c$.  (c) and (d) show the results for the nonlocal hybridization with $V$=0.36 and $t_f=-0.05 t_c$. The black lines in (a) and (c) describe the condition $\mathrm{Im}\Sigma(\omega)/2=\mathrm{max}|V(\bm{k})|$ and the black lines in (b) and (d) describe  the conditions of Eq.(\ref{1}) and $\omega-\mathrm{Re}(h_0) = 0$\label{fig:Sigma}}
\end{figure}
  \begin{figure*}[t]
\includegraphics[width=0.3\linewidth]{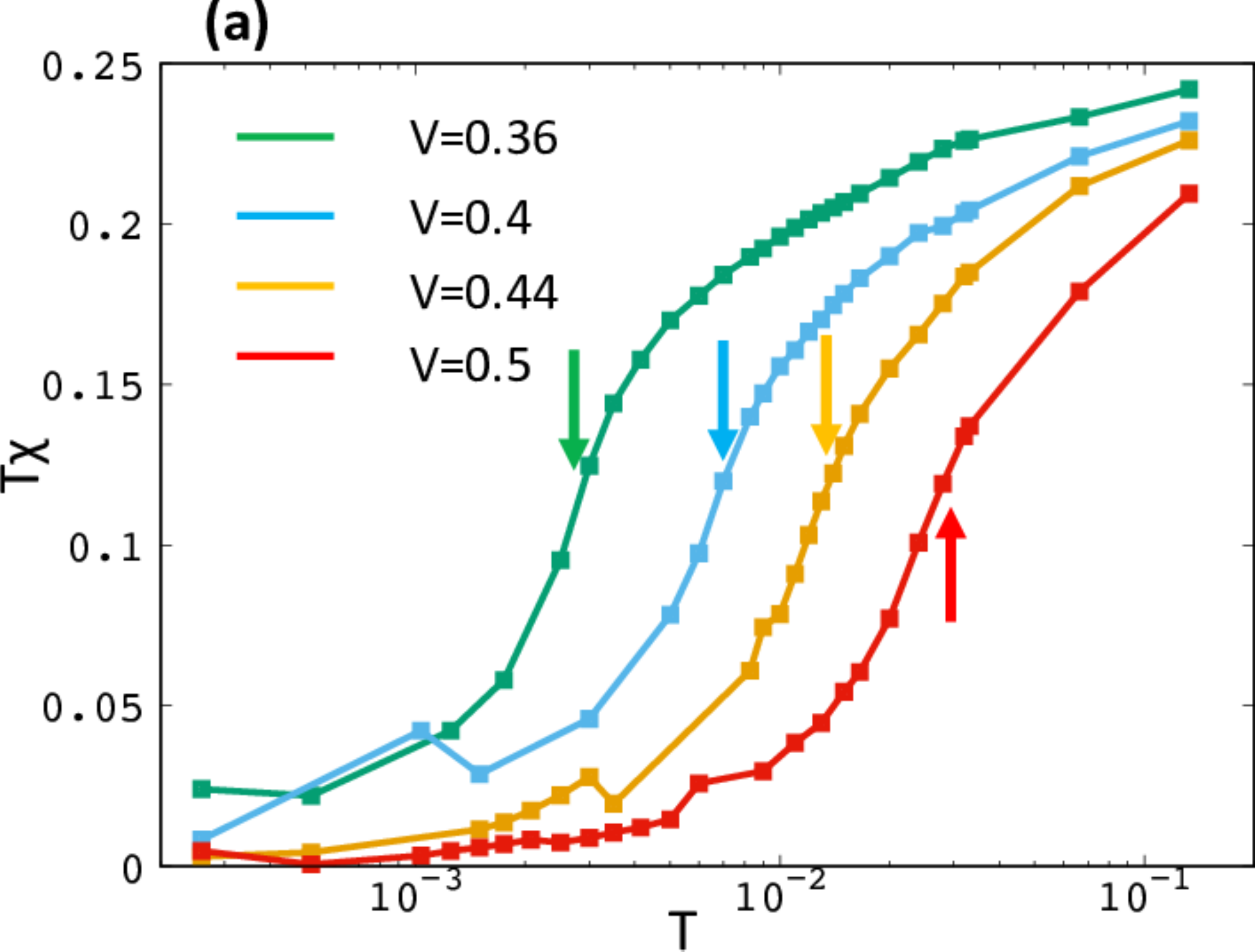}
\includegraphics[width=0.3\linewidth]{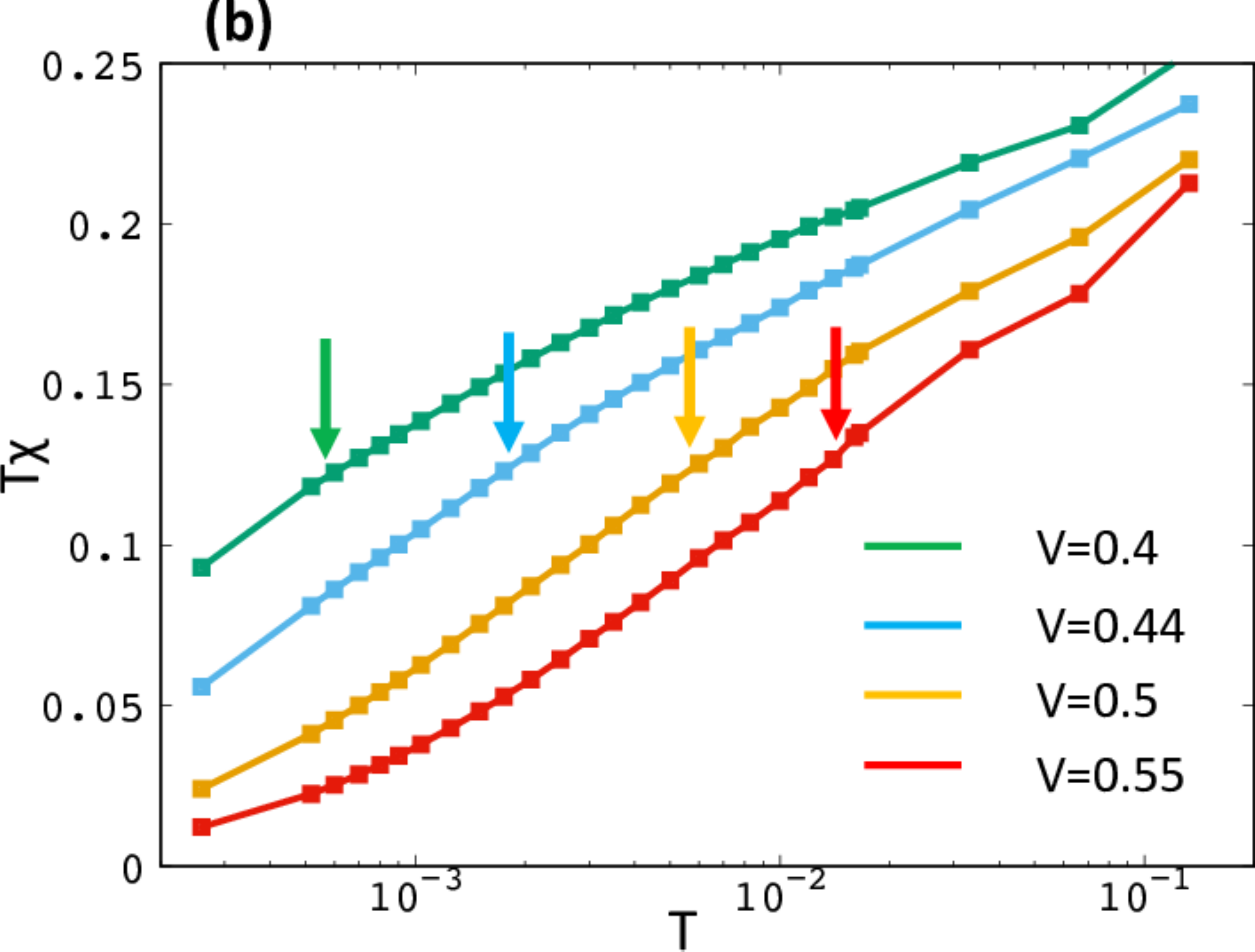}
\includegraphics[width=0.3\linewidth]{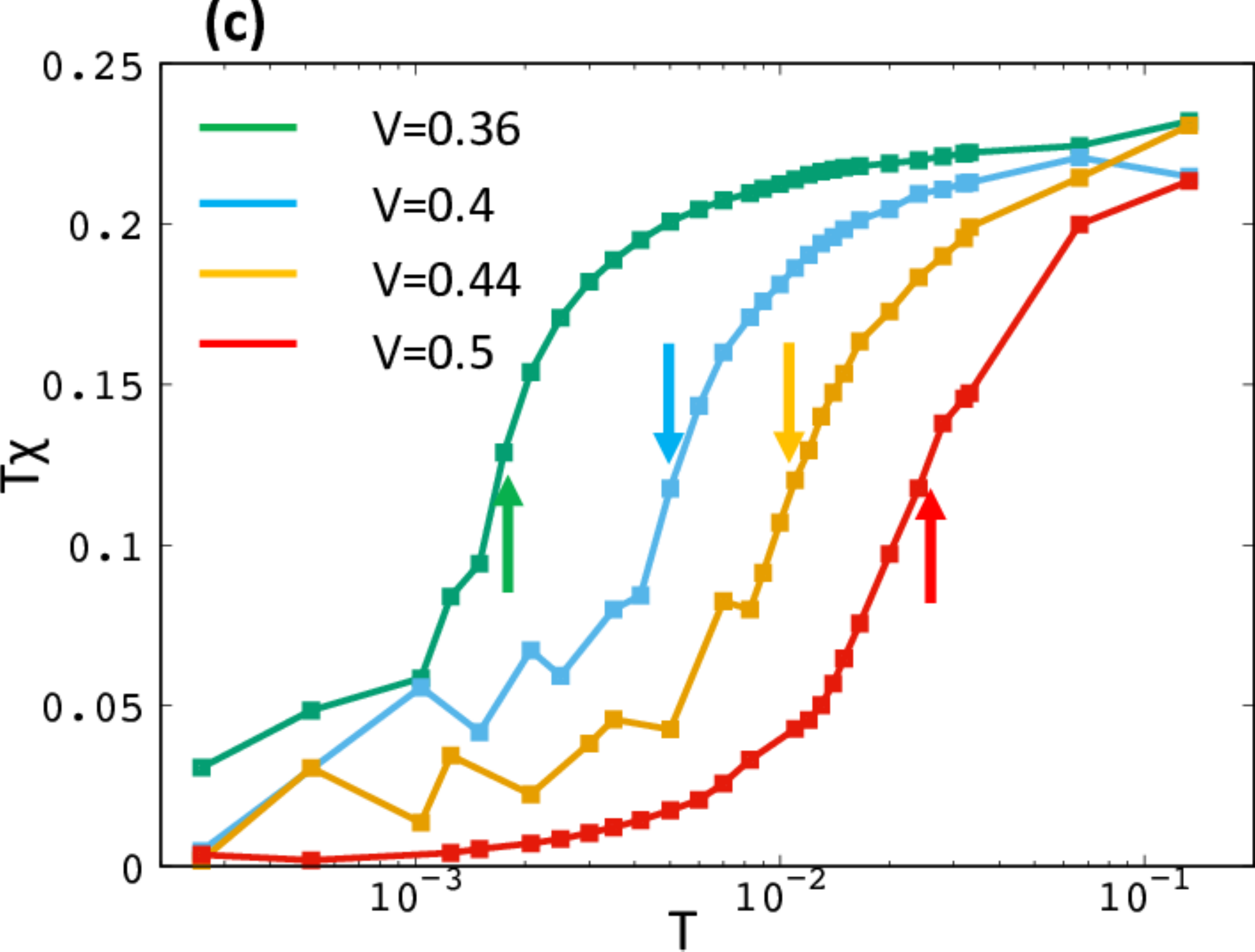}\\

\includegraphics[width=0.3\linewidth]{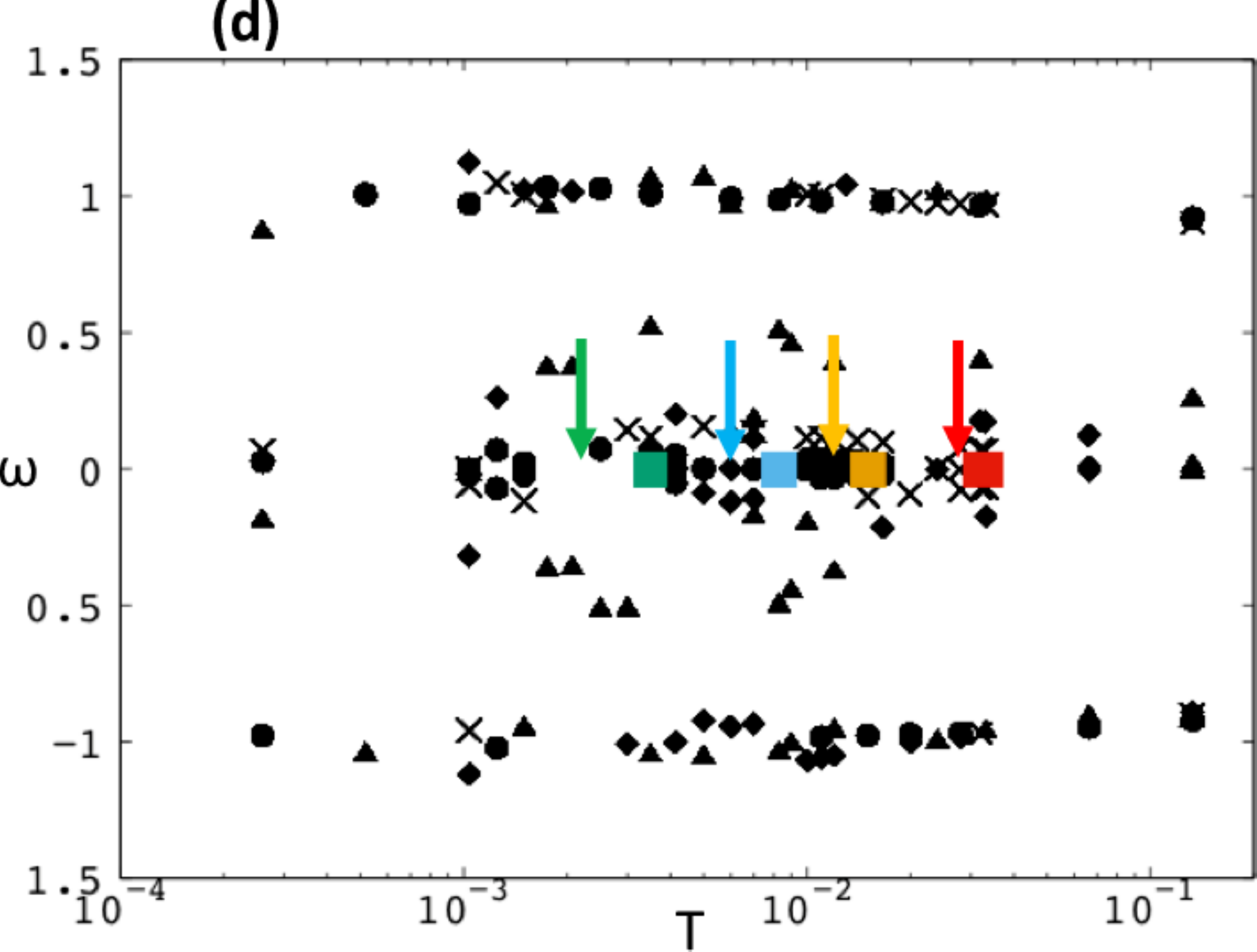}
\includegraphics[width=0.3\linewidth]{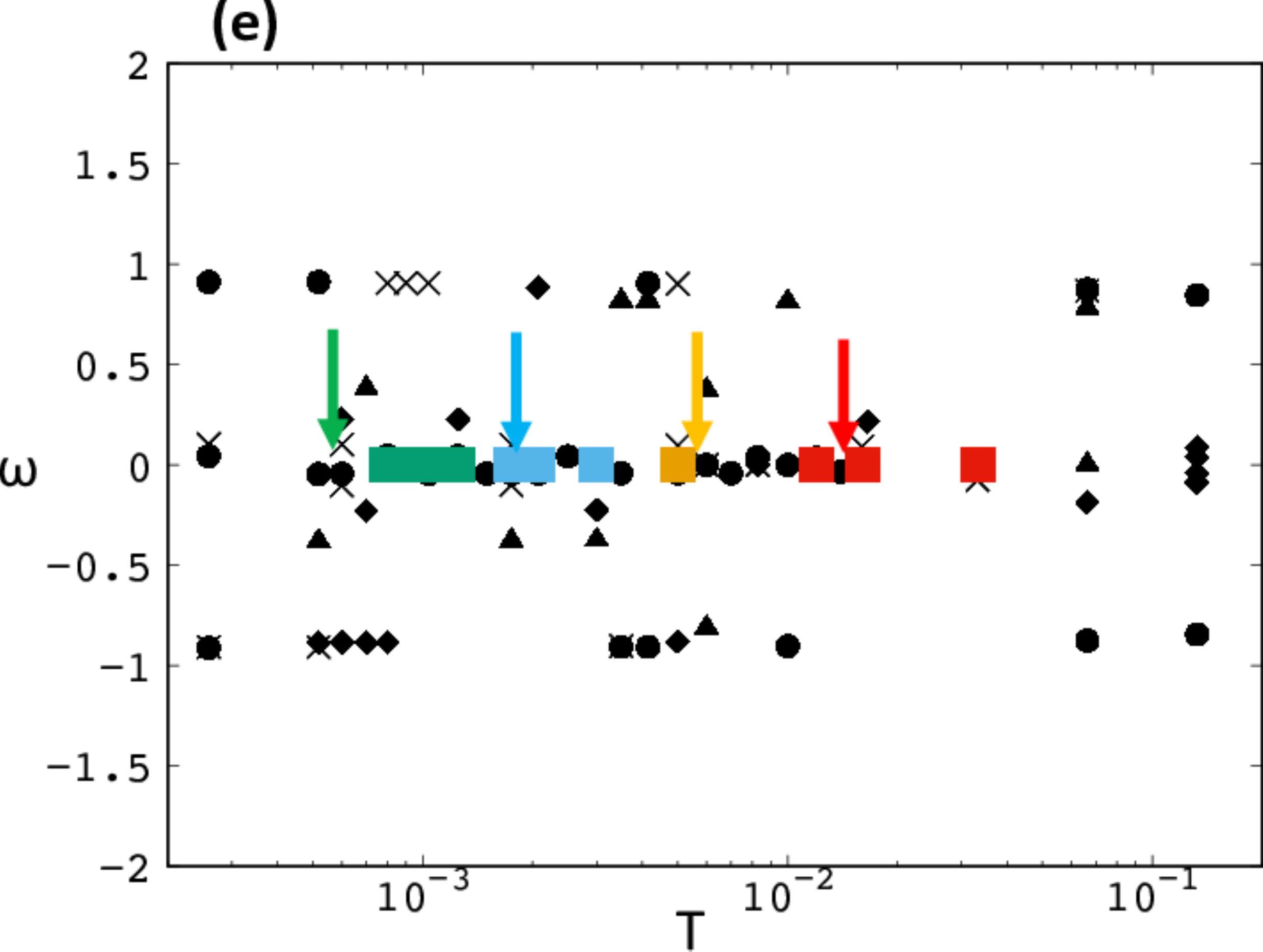}
\includegraphics[width=0.3\linewidth]{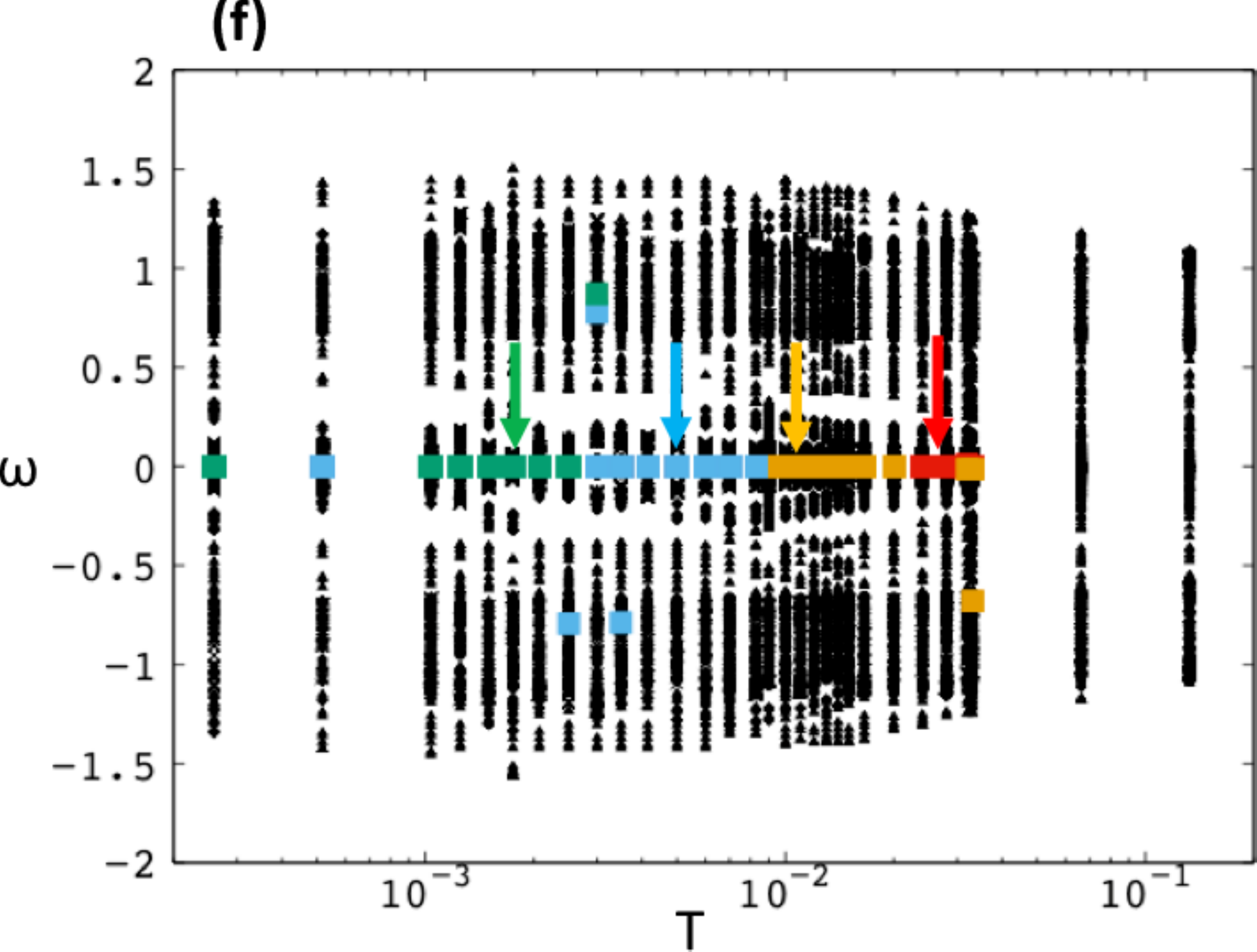}\\

\includegraphics[width=0.3\linewidth]{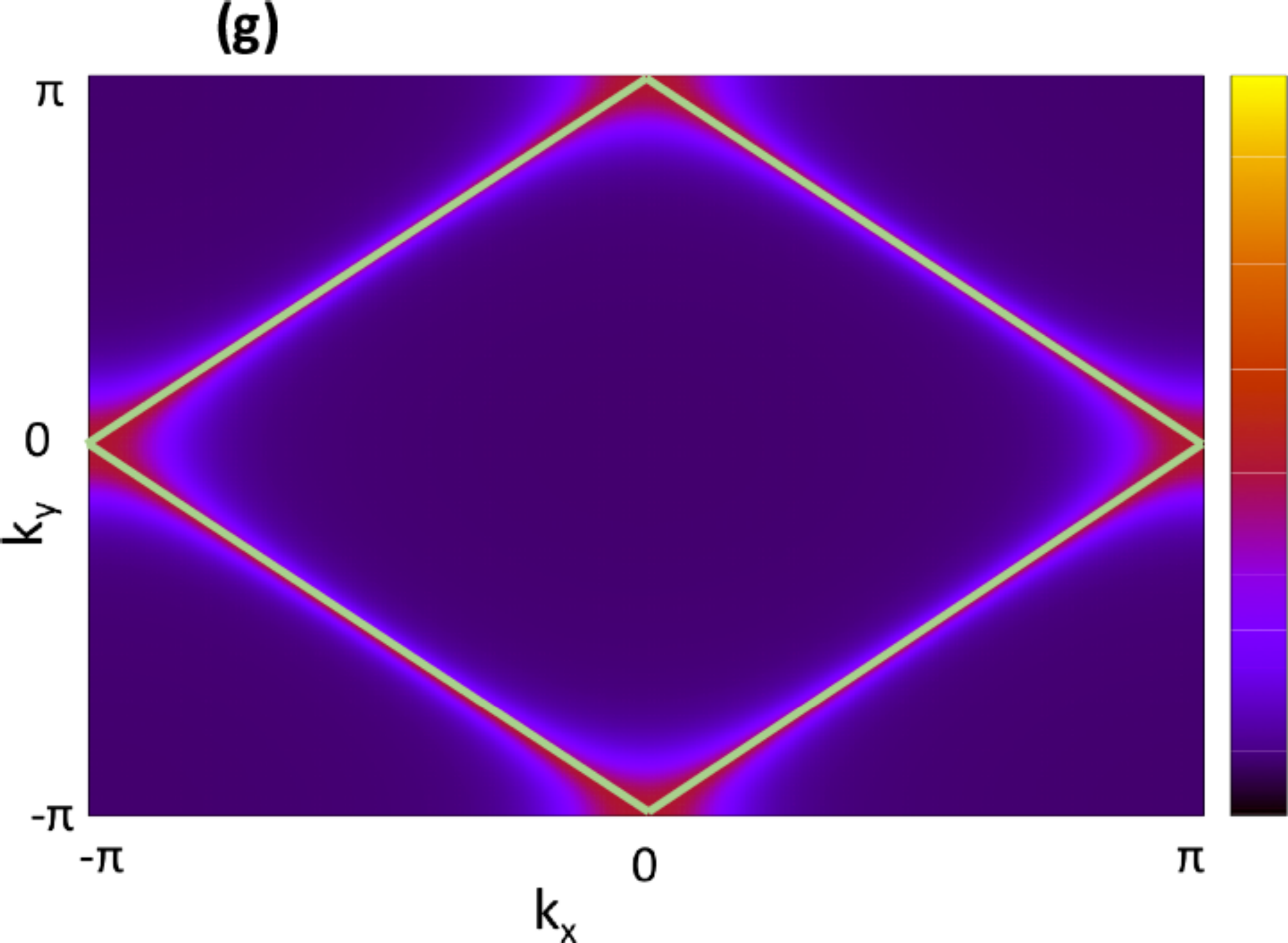}
\includegraphics[width=0.3\linewidth]{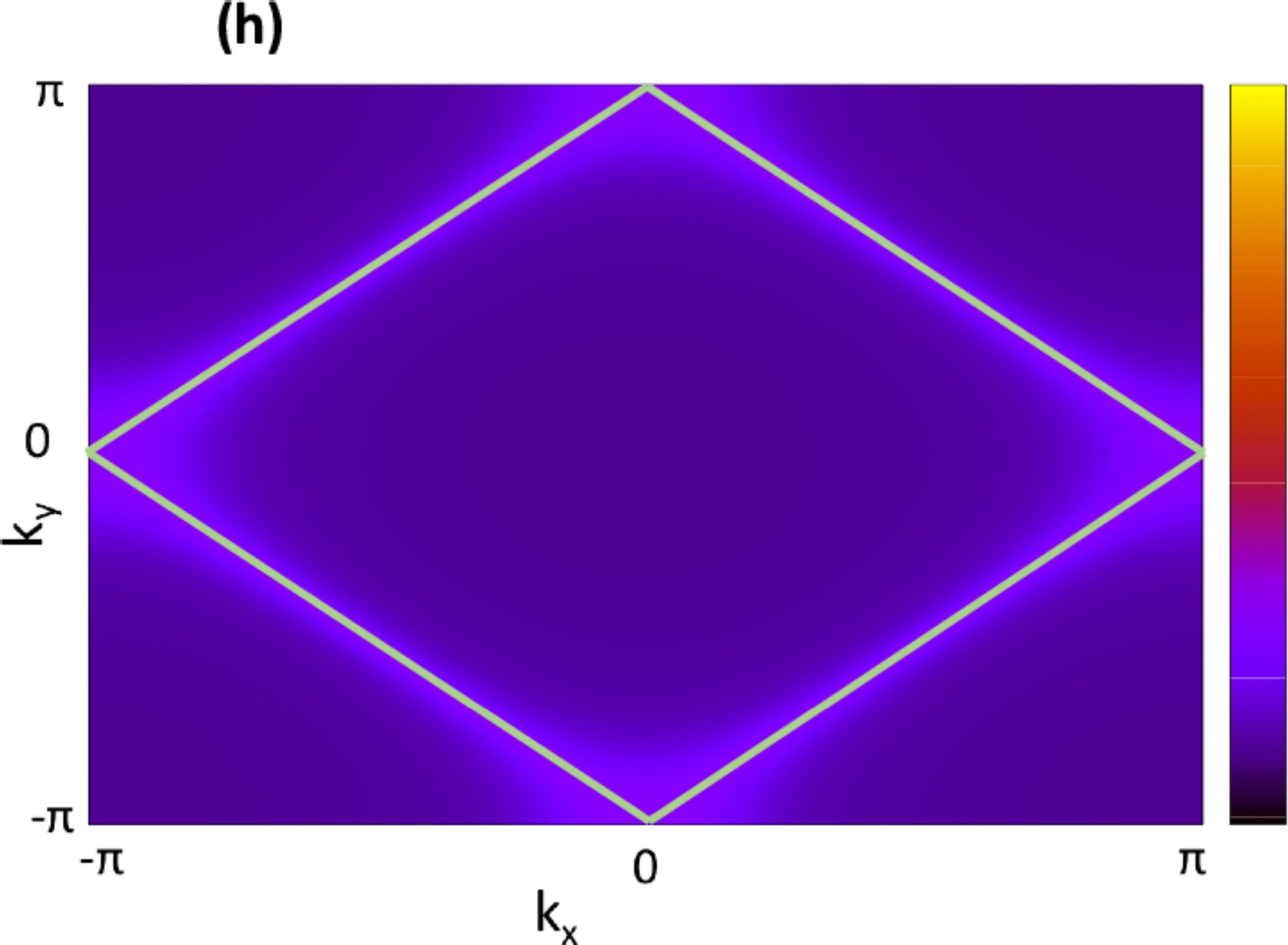}
\includegraphics[width=0.3\linewidth]{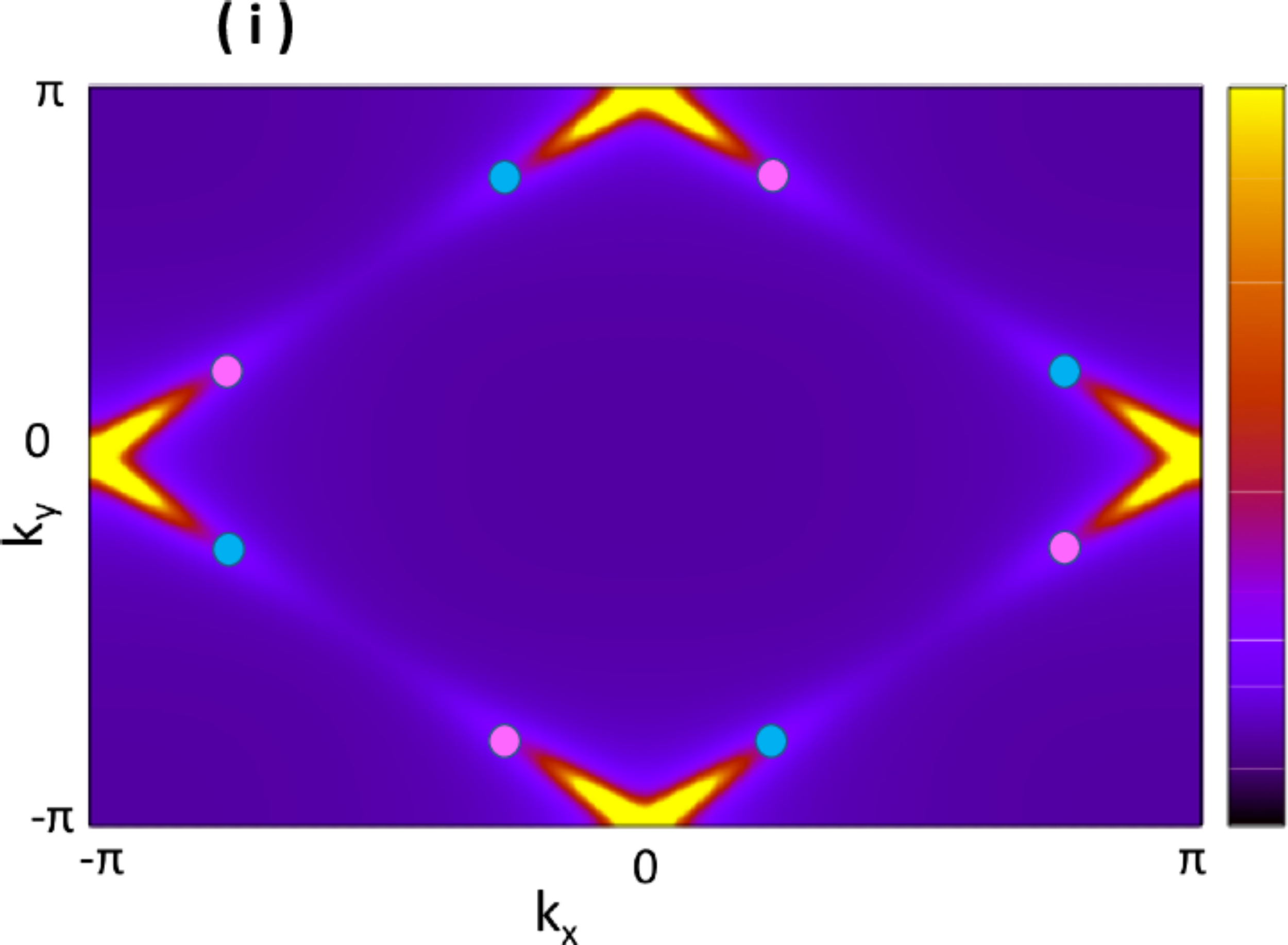}
\caption{  (a)-(f) : Comparison between the Kondo insulator, the heavy-fermion state, and the metallic state with p-wave hybridization for different strengths of $V$. 
(a)-(c) show the local susceptibility. The colors in (a)-(c) correspond to different $V$. The arrows describe the estimated Kondo temperature for each case.
In (a) and (c), the green, skyblue, orange and red plot respectively correspond to $V=0.36, 0.4, 0.44, 0.5$.
In (b), the green, skyblue, orange and red plot respectively correspond to $V=0.4, 0.44, 0.5, 0.55$.
(d)-(f) show the temperature and the frequency dependence of the emergence of the iEPs and the EPs.
The iEPs are drawn as black dots. In (d) and (f), the cross, point, triangle, and square dots are for $V=0.36, 0.4, 0.44, 0.5$.  In (e), the cross, point, triangle, and square dots correspond to $V= 0.4, 0.44, 0.5, 0.55$. 
The EPs are drawn as color plots. We use the same color as in (a)-(c). The arrows describe the Kondo temperature estimated from (a)-(c). 
(g)-(i) show momentum-resolved spectral functions at the Fermi energy around $T_{EP}$.
(g) corresponds to $V$=0.36, $T$=0.0035 shown in (a) and (d). (h) corresponds to $V$=0.4, $T$=0.0007 shown in (b) and (e). (i) corresponds to $V$=0.36,$T$=0.0025 shown in (c) and (f). The parameters are $U$=2, $t_c$=0.8, $t_f$=-0.04, $\mu_c$=0, $\mu_f$=-1.0. 
In (g) and (f), exceptional points form a closed loop in the BZ, highlighted as a green line. In (i), we included exceptional points with vorticity $\pm1/2$ as red and blue points.
\label{fig:Moment2}}
\end{figure*}

In Fig. \ref{fig:Sigma}, we show an example of the temperature dependence of the self-energy calculated by DMFT/NRG.
The model with local hybridization, V=0.36, is shown in Fig. \ref{fig:Sigma}(a) and (b).
Because Eq.(\ref{2}) is independent of the momentum for a system with local hybridization, Eq.(\ref{2}) can be satisfied for all $\bm{k}$ in the BZ and therefore EPs and iEPs can emerge at $\omega$ where the imaginary part of the self-energy crosses the black line.
For the emergence of EPs which have strong spectral weight, additionally $\mathrm{Re}(\omega-h_0)\simeq0$ must be fulfilled.
Because $\epsilon_f/\epsilon_c=$const in our model, the momentum dependence vanishes in Re($w-h_0$)=0. Thus, the condition for the emergence of an EP  can be written as $\epsilon_cw/(\epsilon_c+\epsilon_f)-\mu= \gamma w -\mu =$Re$\Sigma(\omega)$, where $\gamma$ is a constant.
In Fig. 3(b) and (d), this condition is fulfilled when the black line intersects with Re$\Sigma(\omega)$.
We note that even in a model where $\epsilon_f/\epsilon_c$ is not constant, the momentum dependence of Re($\omega-h_0$) is small, because usually  $\epsilon_f$ is much smaller than $\epsilon_c$ .
Thus, we see in Fig. 3 that the condition Re($w-h_0$)=0 can be fulfilled at the Fermi energy.
  
For the system with nonlocal hybridization, shown in Fig. \ref{fig:Sigma}(c) and (d), Eq.(\ref{2}) can be satisfied at $\omega$, where the absolute value of the imaginary part of the self-energy is smaller than the black line, because the strength of the hybridization depends on the momentum. Therefore, EPs and iEPs can appear more easily in this case as we will show in the next section.

\section{Relation between Kondo temperature and $T_{EP}$ in $f$-electron materials }\label{Results}

In Fig. \ref{fig:Moment2} (a)-(c), we show the magnetic moment of the $f$-electrons (contribution of the $f$-electron to the magnetic susceptibility, $T\chi^z_f(T)$). Around the Kondo temperature, the magnetic moment changes from $0.25$ at high temperatures to $0$ at low temperatures, which corresponds to the Kondo screening. 
The magnetic susceptibility in Fig. \ref{fig:Moment2} (a)-(c) is thereby calculated by applying a tiny magnetic field to the system and calculating the induced magnetic polarization of the $f$-electrons. 
We here estimate the Kondo temperature as the temperature where the magnetic moment crosses $T\chi=0.125$ in Fig. \ref{fig:Moment2} (a)-(c) and include an arrow at these temperatures.
We note that this screening is a crossover occurring over a finite temperature range. Thus, the Kondo temperature can also only be determined approximately within the temperarure region where the magnetic moment is screened.
For the metallic system with local hybridization shown in Fig.\ref{fig:Moment2}(b), the screening occurs more slowly compared to the other cases. In this system, the Fermi surface does not vanish below the Kondo temperature so that scattering around the Fermi surface can occur and the imaginary part of the self-energy at the Fermi energy, which prevent the screening of the $f$-electrons, decreases only slowly with lowering the temperature. 
Thus, the screening in Fig. \ref{fig:Moment2}(b) occurs slowly.
Although the system with non-local hybridization shown in Fig. \ref{fig:Moment2}(c) and (f) also has a Fermi surface, it is almost point-like and therefore it induces much less scattering, resulting in a fast screening.

Besides analyzing the Kondo screening, we can use the self-energies obtained by DMFT/NRG, to analyze the emergence of EPs in the spectrum and the temperature at which the EPs appear at the Fermi energy. We show these EPs in Fig. \ref{fig:Moment2} (g)-(i), where we plot iEPs with large eigenvalue $\mathrm{Re}(\omega-h_0)$ as black dots and EPs with $\mathrm{Re}(\omega-h_0)\simeq0$ as colored dots corresponding to the colors of the hybridization shown in Fig. \ref{fig:Moment2} (a)-(c). We see that EPs with  $\mathrm{Re}(\omega-h_0)\simeq0$ appear only in a narrow temperature region for the system with the local hybridization and below a certain temperature for the system with nonlocal hybridization.
In Fig. \ref{fig:Moment2}(e), EPs appear in a wider temperature range than in Fig. \ref{fig:Moment2}(d)
because the self-energy changes only slowly when lowering the temperature. 
Finally, in Fig. \ref{fig:Moment2}(f), EPs can emerge below a certain temperature, because of the momentum dependence of the hybridization, which makes it easier to satisfy Eq. (\ref{2}).
Furthermore, in Fig. \ref{fig:Moment2}(f), EPs appear also far from $\omega=0$. This is possible because, for the system with the nonlocal hybridization, the conditions for the emergence of EPs can be satisfied more easily. 
However, the emergence of EPs far from $\omega=0$ seems to be irrelevant to the Kondo effect because the Kondo effect stems from the scattering around the Fermi surface.

Comparing the temperature at which EPs with $\mathrm{Re}(\omega-h_0)\simeq0$ appear and the temperature in which the magnetic moment of the $f$-electron is screened, we see that both temperatures match very well.
Thus, we conclude that the Kondo temperature is closely related to the temperature where EPs emerge at the Fermi energy. 
When lowering the temperature, the self-energy at the Fermi energy changes very strongly around the Kondo temperature, which results in a situation in which Eq. (\ref{1}) and (\ref{2}) can be easily fulfilled at the Fermi energy.
For the system with the $p$-wave hybridization, the EPs start to emerge at the Kondo temperature when the absolute value of the imaginary part of the self energy becomes smaller than the hybridization strength.

In Fig. \ref{fig:Moment2} (d)-(f), we can also see many iEPs with large $\mathrm{Re}(\omega-h_0)$, which appear at almost all temperatures. These iEPs are mainly related to the imaginary part of the self-energy away from the Fermi energy, particularly in the Hubbard bands, which are nearly temperature independent and are irrelevant to the Kondo effect.

 Figures \ref{fig:Moment2} (g)-(i) show the Fermi surface at the temperature at which EPs appear at the Fermi energy for the systems with local hybridization and for the system with $p$-wave hybridization.
 For the system with the local hybridization, EPs appear only at $T_{EP}$ and form a closed loop in the BZ.
We note that it is not possible to define a vorticity at $\omega=0$ for this closed loop, which is different from the symmetry-protected
exceptional ring in systems with chiral symmetry\cite{PhysRevB.99.041406,PhysRevB.99.041202,PhysRevB.99.121101,yoshida2019exceptional}. Therefore, we believe that this loop of exceptional points will change into isolated EPs connected by bulk Fermi arcs when taking into account a momentum-dependent self-energy.
For the system with the $p$-wave hybridization, EPs appear as isolated points at the Fermi energy in the spectrum and have nonzero vorticity, and are thus topologically protected. 
These EPs change their position in the BZ satisfying Eq.(\ref{1}) when changing the temperature, and finally merge and disappear at zero temperature.

\section{Extension of the exceptional manifolds to $\omega$-space }\label{EPR}
For the emergence of exceptional points, two equations (Eq.(\ref{1}) and Eq.(\ref{2})) must be satisfied. Thus, in a d-dimensional model (2-dimensional momentum space in this paper), exceptional points will generally form (d-2) dimensional manifolds. The dimension of the exceptional manifold might be higher, if additional symmetries do exist. For example, in 2D systems with chiral symmetry\cite{PhysRevB.99.121101}, one of the two conditions for EPs, such as Eq. (\ref{1}) and (\ref{2}), is always satisfied which leads to (d-1) dimensional exceptional manifolds.
  \begin{figure*}[t]
\includegraphics[width=0.45\linewidth]{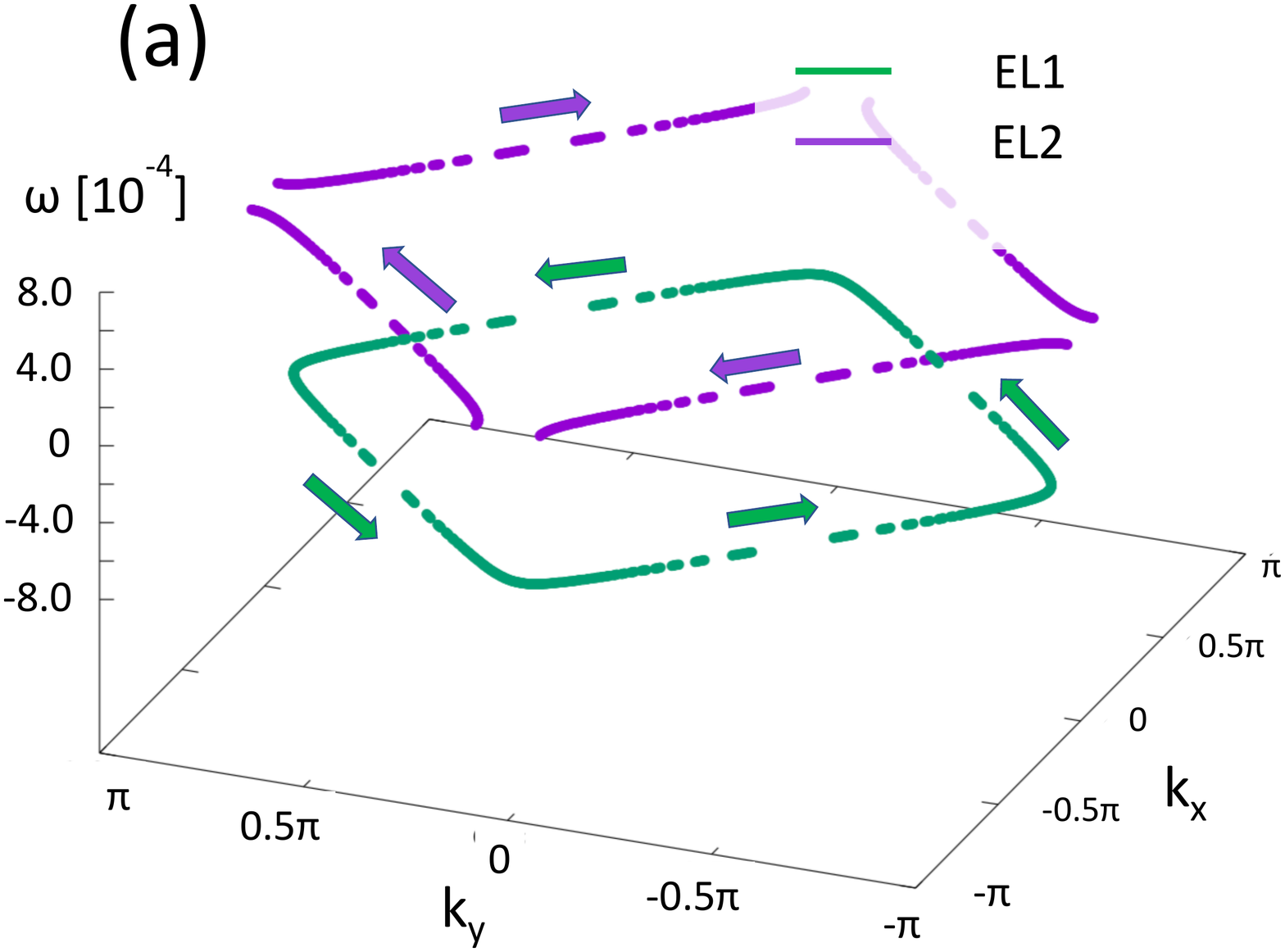}
\includegraphics[width=0.45\linewidth]{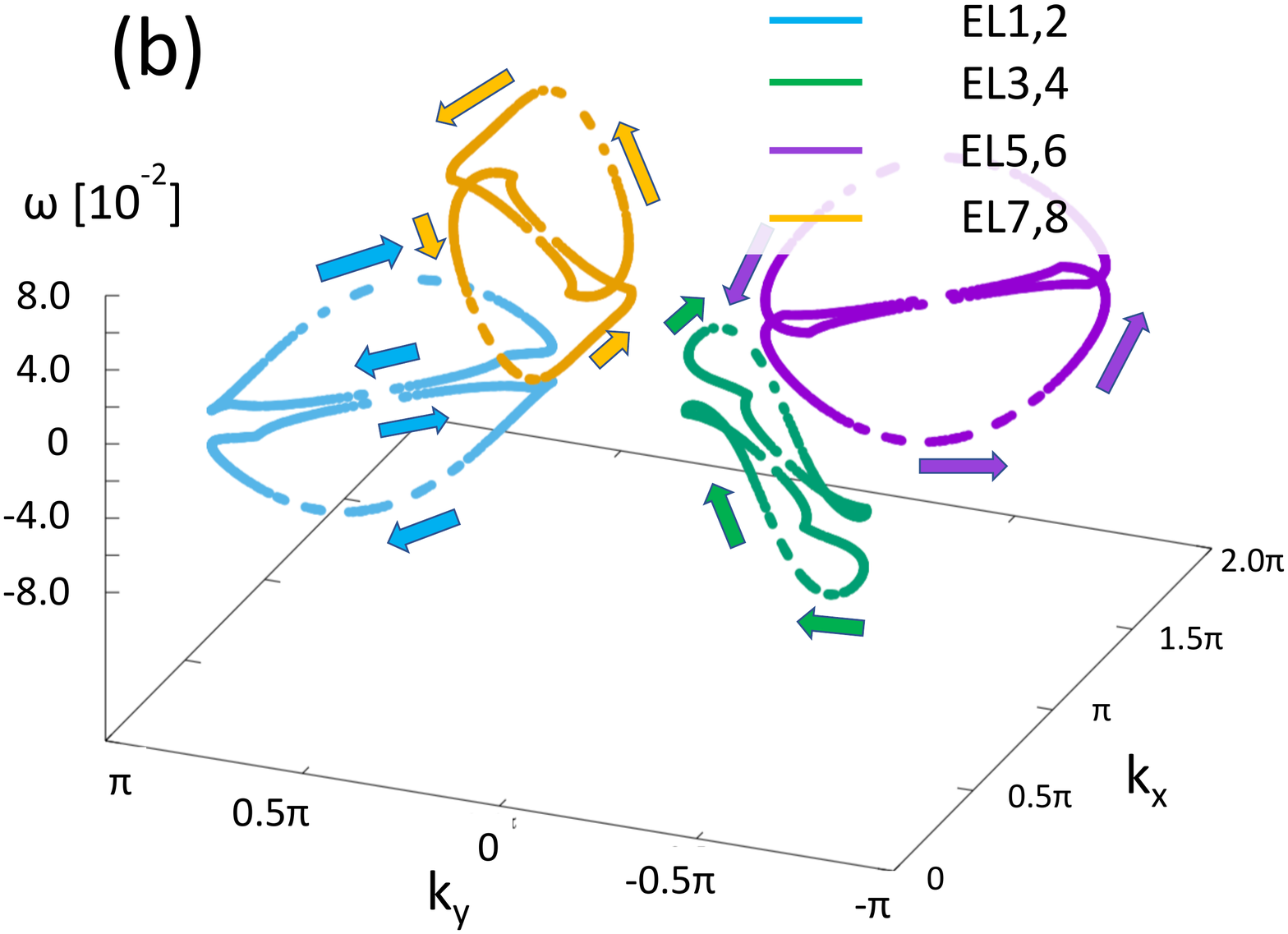}\\
\includegraphics[width=0.45\linewidth]{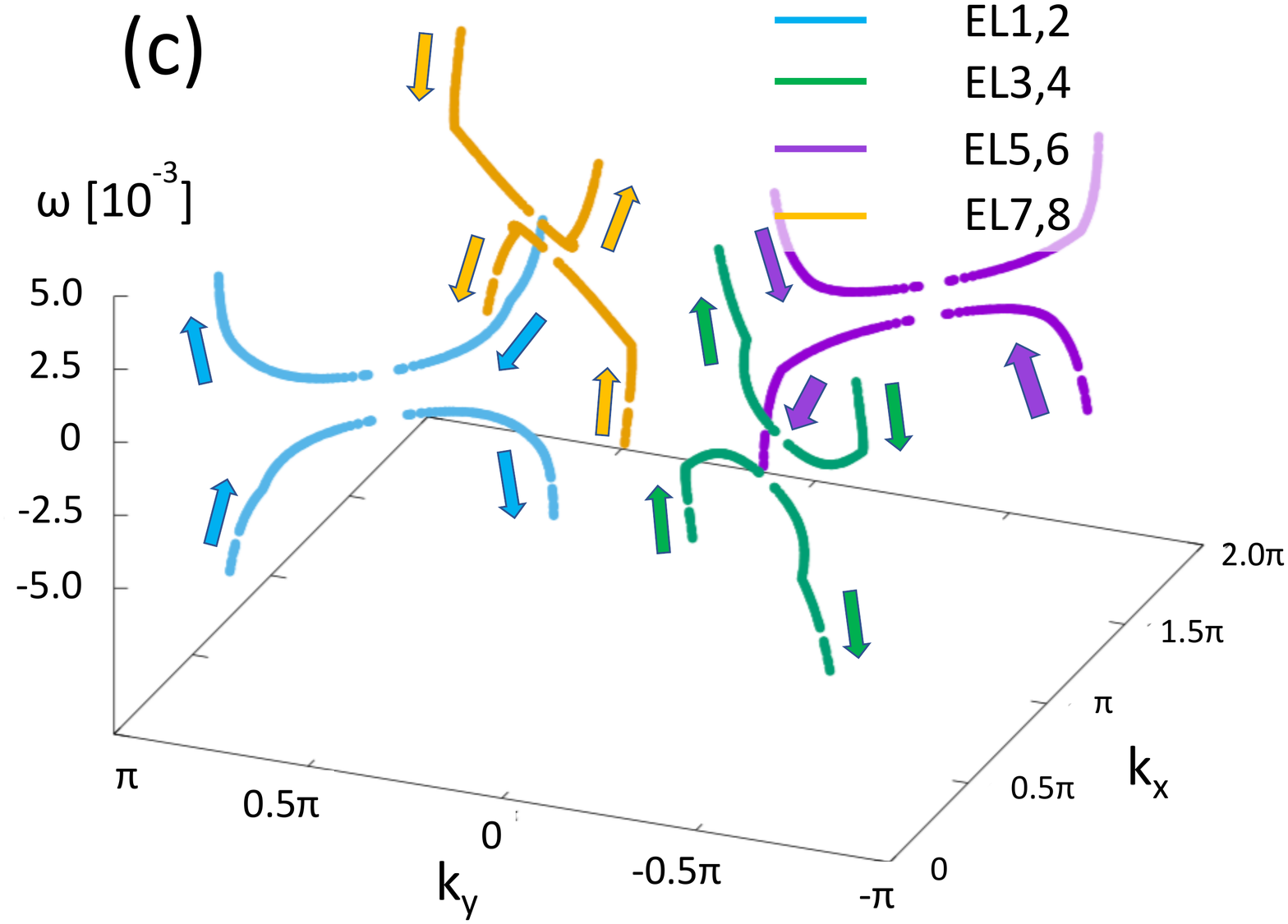}
\includegraphics[width=0.45\linewidth]{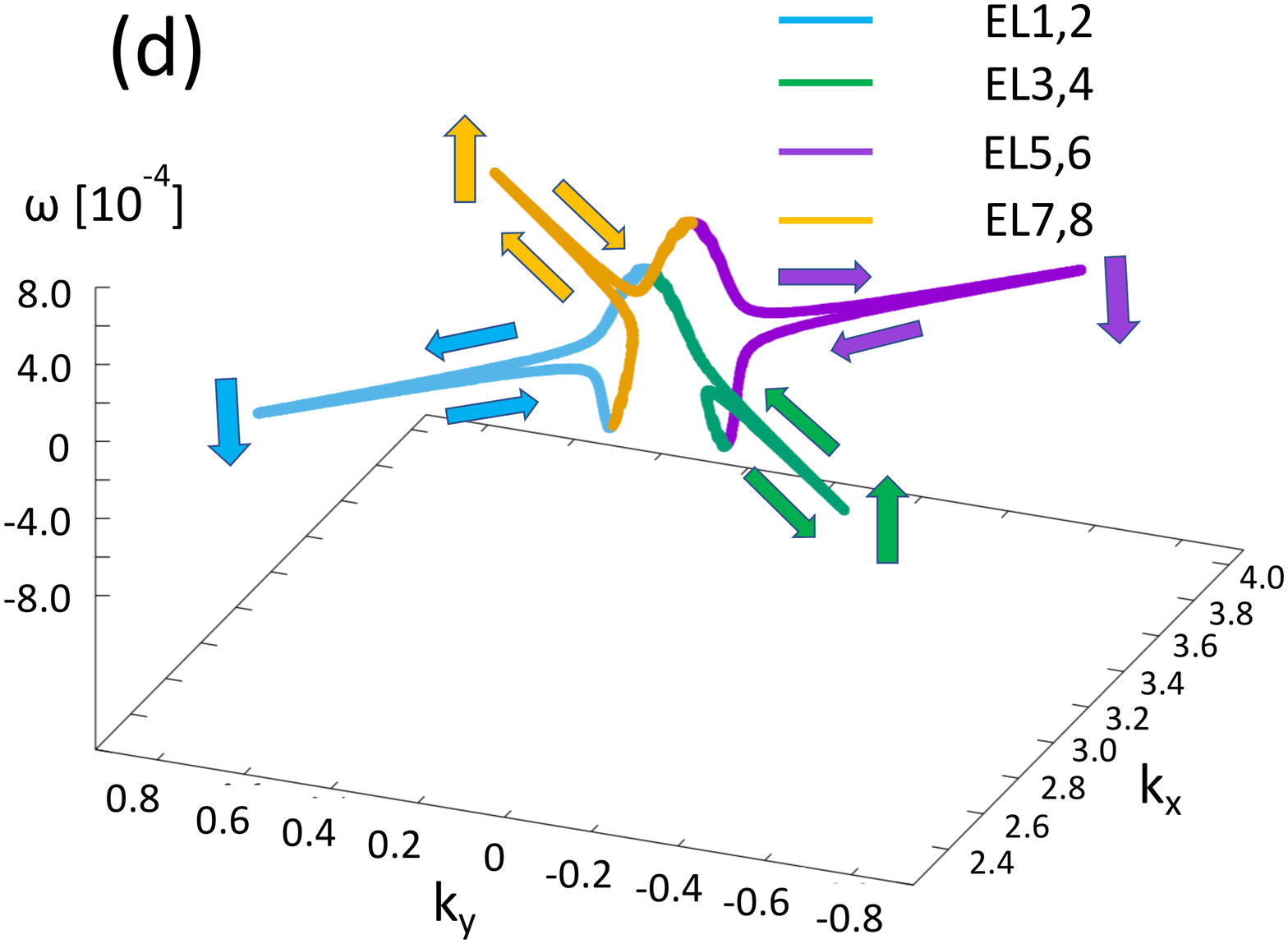}
\caption{Exceptional points of $\mathcal{H}_{eff}(k_x,k_y,\omega)$ in the system with local hybridization for (a) and with nonlocal hybridization for (b)-(d). The parameters are $V_l=0.4$,$t_f/t_c=-0.05$,$T=0.016$ in (a),$V_p=0.4$,$t_f/t_c=-0.05$,$T=0.016$ in (b) and (c), and $V_l=0.4$,$t_f/t_c=-0.05$,$T=0.002$ in (d). (c) is a magnification of the exceptional manifold shown in (b) close to the Fermi energy. 
The vorticities in (a) are calculated by Eq.(\ref{Ext}).\label{fig:EPR}}
\end{figure*}
Besides the spatial dimension, the effective Hamiltonian in strongly-correlated systems also depends on the frequency, $\omega$, because the self-energy depends on the frequency.
Thus, the inclusion of frequency will increase the dimension of the exceptional manifolds. Previous studies have only focused on the Fermi energy, ignoring the frequency dependence of the exceptional manifolds.
We note that a frequency dependent effective Hamiltonian occurs in situations when focusing on a subsystem and integrating out the rest of the total system, even though the full system is described by a frequency-independent Hamiltonian\cite{FESHBACH1958357,FESHBACH1962287,breuer2002theory}. Because we here focus on the one-particle Green function, the effective Hamiltonian depends on the frequency. 

In Fig. \ref{fig:EPR}, we show the exceptional manifolds for the local and the nonlocal hybridization and different temperatures in the ($\bm{k},\omega$)-space. Until now, we have focused only on exceptional points close to the Fermi energy. Fig. 5(a) shows the exceptional loops in the system with local hybridization for a temperature above the Kondo temperature. As described above, in the system with local hybridization, Eq.(\ref{1}) and Eq.(\ref{2}) do not depend on the momentum and thus exceptional manifolds are loops in the momentum space. At temperatures above the Kondo temperature, we find one loop above the Fermi energy and one loop below the Fermi energy.
Lowering the temperature towards the Kondo temperarure, these exceptional loops move towards the Fermi energy. At the Kondo temperature, these exceptional loops merge at the Fermi energy.

We note that by extending our considerations to the ($\bm{k},\omega$)-space, we are able to define and calculate the vorticity of these loops by
\begin{align}
 &\nu=\oint_{\mathrm{EP}} \frac{d\bm{k}^{\prime}}{2\pi i} \cdot \bm{\nabla}_{\bm{k}^{\prime}} \mathrm{log} \  \mathrm{det} \mathcal{H}_{eff} (\bm{k},\omega),\label{Ext}
\end{align}
where $\bm{k}^{\prime}$ is defined on the plane which is perpendicular to the tangent vector of the exceptional loop. The line integral is done in mathematical positive direction. We then define the direction of the exceptional loops, shown in Fig. \ref{fig:EPR}, so that the vorticities defined in Eq.(\ref{Ext}) become 1/2.
Further details are explained in Appendix \ref{app:Def}.
We note that when considering a momentum dependent self-energy, these loops become distorted. Thus, looking at the Fermi energy, the exceptional manifold will appear as points.
  
Fig. 5(b)-(d) show the exceptional manifolds for the system with nonlocal hybridization. 
Because Eq. (\ref{1}) and Eq.(\ref{2}) depend on the momentum for a nonlocal hybridization, the exceptional manifolds are points in the momentum space for fixed $\omega$. Fig. \ref{fig:EPR}(b) and (c) show these exceptional points in ($\bm{k},\omega$)-space at high temperatures. Including the $\omega$-space, these exceptional points form closed loops. 
Furthermore, we show in Fig. \ref{fig:EPR}(c) a magnification of (b) around the Fermi energy, which demonstrates the absence of exceptional points at the Fermi energy for high temperatures. 
When lowering the temperature, one closed loop of EPs is formed above $\omega=0$ and one closed loop is formed below $\omega=0$. At the Kondo temperature, these loops touch and merge at $\omega=0$ to a single 1D exceptional manifold  as shown in Fig. \ref{fig:EPR}(d).

We conclude that the crossover from localized to itinerant $f$-electrons in the system with local hybridization goes hand in hand with the merging and vanishing of two exceptional loops at the Fermi energy. In the system with nonlocal hybridization, different exceptional loops in the ($\bm{k},\omega$)-space change their topology at the Kondo temperature and generate EPs at the Fermi energy.

\section{Conclusion and Discussion}\label{Discuss}
In summary, we have shown a relation between the Kondo temperature and the emergence of  exceptional points at the Fermi energy in $f$-electron materials.
Particularly, we have studied the Kondo insulator and the metallic state with a local hybridization, and the semimetallic state with a $p$-wave hybridization in the 2D periodic Anderson model by DMFT/NRG. 
Around the Kondo temperature, $f$-electrons change from localized to itinerant when lowering the temperature.
Thus, the emergence of EPs at the Fermi surface is a sign for the crossover between localized and itinerant $f$-electrons.
We have also shown that the exceptional manifolds have a one more higher dimensional structure when considering the $\omega$-dependence of the effective Hamiltonian.
For the system with the local hybridization, there are exceptional loops above and below $\omega=0$ in the ($\bm{k},\omega$)-space at high temperature, which merge and disappear around the Kondo temperature.  
For the $p$-wave hybridization, there are four exceptional loops each above and below $\omega=0$ in the ($\bm{k},\omega$)-space at high temperature, which merge and change their topology around the Kondo temperature. 
Contrary to the system with local hybridiation, these exceptional manifolds at the Fermi energy are stable for a wide range of temperatures below the Kondo temperature and these EPs are connected by bulk Fermi arcs. 
  
We can naturally expect that the relation between the emergence of exceptional points at the Fermi energy and the Kondo temperature hold for three-dimensional systems because the DMFT results become more accurate for higher dimensions. Because the band structure of the 3D system can be understood by stacking 2D systems, the 3D f-electron material should host robust exceptional rings in the case of a momentum dependent hybridization and exceptional surfaces in the case of a local hybridization at the Fermi energy.

\begin{acknowledgments}
 This work is partly supported by JSPS KAKENHI Grants No. JP18H04316, No. JP18H05842 and No. JP18K0351.
Computer simulations have been performed on the supercomputer of the ISSP in the University of Tokyo.
\end{acknowledgments}

\appendix
\section{definition of the vorticity in ($\bm{k},\omega$)-space}\label{app:Def}
\begin{figure}[b]
\includegraphics[width=0.8\linewidth]{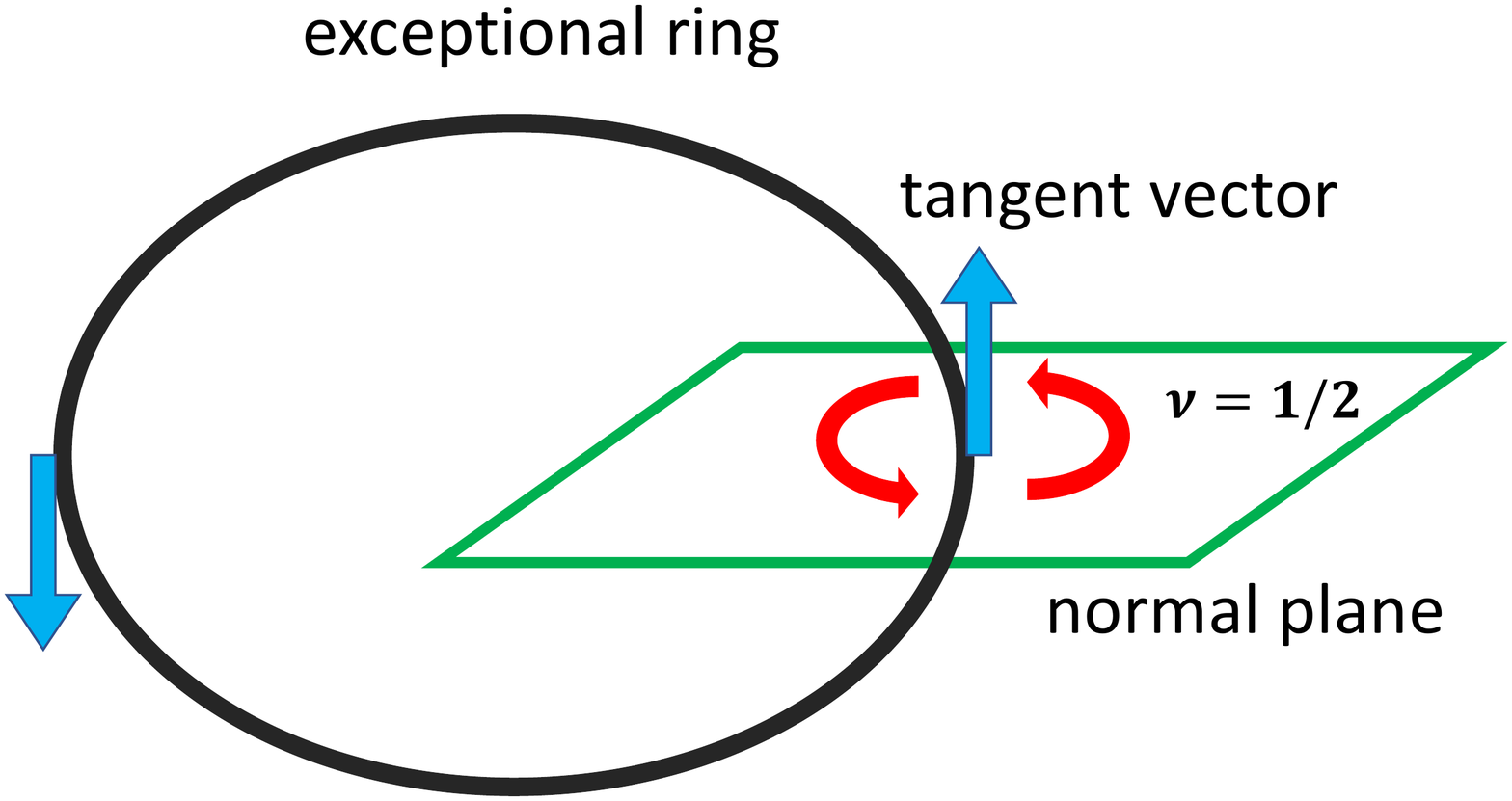}
\caption{Visualization of the tangent vector, normal plane, and the vorticity.}\label{fig:Def_Vor}
\end{figure}
Here, we explain the details of the definition of the vorticity in the ($\bm{k},\omega$)-space.
In (2+1) dimensional systems, exceptional points form a loop in the ($\bm{k},\omega$)-space. In order to define the vorticity of the exceptional loop, we set up the plane in the ($\bm{k},\omega$)-space to which the tangent vector of the loop is the normal vector ( see Fig. \ref{fig:Def_Vor}).  We calculate the vorticity by doing the following line integral in mathematical positive direction,
\begin{align}
 &\nu=\oint_{\mathrm{EP}} \frac{d\bm{k}^{\prime}}{2\pi i} \cdot \bm{\nabla}_{\bm{k}^{\prime}} \mathrm{log} \  \mathrm{det} \mathcal{H}_{eff} (\bm{k},\omega),\label{Ext2}
\end{align}
where $\bm{k}^{\prime}$ is defined on the plane as shown in Fig.\ref{fig:Def_Vor}.

However, the sign of the integral still depends on the orientation of the plane, which depends on the direction of the tangent vector. Therefore, we choose the direction of the tangent vector so that the integral in Eq. (\ref{Ext2}) becomes 1/2. This uniquely defines a direction for the exceptional loop in the ($\bm{k},\omega$)-space, which is shown in Fig. \ref{fig:EPR}. 
To define the vorticity in this way, we need to define a plane perpendicular to the exceptional manifold. Thus, it is necessary that the dimension of the exceptional manifold is ($d-2$), where d is the dimension of the ($\bm{k},\omega$)-space.
By using Eq.(\ref{1}) and (\ref{2}), we see that exceptional points can merge and disappear only if their tangent vectors are antiparallel, when two loops touch.
We note that in systems whose Hamiltonian is described by a larger than 2 by 2 matrix, e.g. a system including more than two orbitals, exceptional points generated by different pairs of eigenvalues can touch with arbitrary direction without merging. 
Furthermore, we note that a similar definition might be used to analyze $2$-dimensional exceptional manifolds in ($3+1$)-dimensional systems.

\bibliography{paper}

\end{document}